\documentclass{ws-ijmpe}
\usepackage[super,compress]{cite}

\usepackage[english]{babel}
\usepackage[utf8x]{inputenc}

\usepackage{latexsym}
\usepackage{amsmath}
\usepackage{amssymb}

\usepackage{supertabular} 
\usepackage{placeins}
\usepackage{epsfig}
\usepackage{graphicx}
\usepackage{hyperref} 

\begin{document}

\markboth{J. Segovia, D.R. Entem, F. Fernandez and E. Hernandez}{Constituent Quark Model
Description of Heavy Meson Decays}

\catchline{}{}{}{}{}
%%%%%%%%%%%%%%%%%%%%%%%%%%%%%%%%%%%%%%%%%%%%%%%%%%%%%%%%%%%%%%%%%%%%

\title{CONSTITUENT QUARK MODEL DESCRIPTION OF CHARMONIUM PHENOMENOLOGY}

\author{J. SEGOVIA}

\address{Physics Division, Argonne National Laboratory, 9700 South Cass Avenue \\
Argonne, Illinois 60439-4832, United States of America \\ jsegovia@anl.gov}

\author{D.R. ENTEM}

\address{Grupo de F\'{\i}sica Nuclear e Instituto Universitario de
Física Fundamental y Matemáticas, Universidad de Salamanca, Casas del Parque S/N \\
Salamanca, E-37008, Spain \\ entem@usal.es}

\author{F. FERNANDEZ}

\address{Grupo de F\'{\i}sica Nuclear e Instituto Universitario de
Física Fundamental y Matemáticas, Universidad de Salamanca, Casas del Parque S/N \\
Salamanca, E-37008, Spain \\ fdz@usal.es}

\author{E. HERNANDEZ}

\address{Grupo de F\'{\i}sica Nuclear e Instituto Universitario de
Física Fundamental y Matemáticas, Universidad de Salamanca, Plaza de la Merced S/N \\
Salamanca, E-37008, Spain \\ gajatee@usal.es}

\maketitle

\begin{history}
\received{\today}
%\revised{Day Month Year}
%\accepted{Day Month Year}
%\comby{(xxxxxxxxxx)}
\end{history}

\begin{abstract}
We review how quark models are able to describe the phenomenology of the  charm
meson sector. The spectroscopy and decays of charmonium and open charm mesons are
described in a particular quark model and compared with the data and the results of
other existing models in the literature. A quite reasonable global description of
the heavy meson spectra is reached. A new assignment of the $\psi(4415)$ resonance as
a $3D$ state leaving aside the $4S$ state to the $X(4360)$ is tested through the
analysis of the resonance structure in $e^{+}e^{-}$ exclusive reactions around the
$\psi(4415)$ energy region. We make tentative assignments of some of the $XYZ$
mesons.
To elucidate the structure of the $1^{+}$ $c\bar{s}$ states, i.e. $D_{s1}(2460)$ and
$D_{s1}(2536)$, we study the strong decay properties of the $D_{s1}(2536)$ meson. 
We also perform a calculation of the branching fractions for the semileptonic decays
of $B$ and $B_{s}$ mesons into final states containing orbitally excited charmed and
charmed-strange mesons, which have become a very important source of information
about the structure of heavy mesons. Analysis of the nonleptonic $B$ meson decays
into $D^{(\ast)}D_{sJ}$ are also included.
\end{abstract}

\keywords{potential models; Heavy quarkonia; charmed mesons; bottom mesons; models of
strong interactions; leptonic and semileptonic decays.}

\ccode{12.39.Pn, 14.40.Pq, 14.40.Nd, 14.40.Lb, 12.40.-y, 13.20.-v}

%\tableofcontents

%%%%%%%%%%%%%%%%%%%%%%%%%%%%%%%%%%%%%%%%%%%%%%%%%%%%%%%%%%%%%%%%%%%%%%%%%%%%%%%%
\FloatBarrier
%%%%%%%%%%%%%%%%%%%%%%%%%%%%%%%%%%%%%%%%%%%%%%%%%%%%%%%%%%%%%%%%%%%%%%%%%%%%%%%%

\section{Introduction}

After the November revolution in 1974, the next significant step in the understanding of charmonium physics was the starting of
the experimental activity at the dawn of the XXI century of the B-factories and the advent of the LHCb.

These machines have produced a huge amount of data which will allow a better understanding of the quarkonium phenomenology.
 Reviews of the theoretical importance and experimental
status of heavy quarkonium have recently been given, among others, by
Quigg~\cite{quigg2004quarkonium}, Galik~\cite{galik2004quarkonium}, the CERN
Quarkonium Working Group~\cite{brambilla2004heavy},
Seth~\cite{seth2005new,seth2005heavy,seth2006heavy}, and
Swarnicki~\cite{skwarnicki2005cleo}.

Although the spectrum of charmonium and their transitions have been the subject of a great number of studies, from the seminal paper of 
Eichten~\cite{PhysRevD.17.3090}
to the more recent  results presented by Radford and Repko~\cite{PhysRevD.75.074031},
there are only few global studies of all the data produced at the B factories. As most of the new states lie above the open charm 
threshold, the study of its decay properties through the $D^{(*)}\bar D^{(*)}$ channels are very important. This implies a consistent 
description of both,
the parent meson and the $D^{(*)}$ mesons involved in the decay. Furthermore, most of the new resonances are produced through a weak 
process, so that a wealth of information 
about the new states can be obtained from the study of the weak decay of B mesons.

All these processes have been partially studied in the literature but the aim of this review is to present a coherent description of as many experimental data as possible in an unique framework. We will provide the comparison with the results of other models where there exist.

With this idea, after a short introduction to the model used, we will describe the spectroscopy of the hidden and open charm mesons as $q\bar q$ 
states, discussing whose of the new XYZ states can be assigned to this structure. The description of theses states  will be complemented with the study of the electromagnetic and strong decays.

One of the outcoming of our calculation is a new quantum number assignment of the $\psi(4415)$ state motivated by the appearance in the $J^{PC}=1^{--}$ spectrum of the $X(4360)$ state.  A detailed study of the reactions $e^+e^-\rightarrow D^0D^-\pi^+$ and $e^+e^-\rightarrow D^0D^{*-}\pi^+$ has been performed in order to justify our result.

In the same way, a study of the decay properties of the $D_{s_1}(2536)^+$ meson has been performed 
and compare with the data of the BELLE collaboration to give more insight in the structure of the $1^+$ $c\bar s$ states.

Finally, most of the new states have been discovered from the semileptonic and non-leptonic B decays into open charm states. 
The theoretical analysis of the data, which includes both weak and strong processes, opens an interesting possibility to study 
the structure of this type of mesons.

The paper is organized as follows. In Sec.~\ref{sec:ConQuaMod} we will introduce our
constituent quark model, paying special attention to the terms that determine the
spectra of heavy mesons. After that, we will present in Sec.~\ref{sec:Spectroscopy}
the spectrum of  hidden-charm and open charm mesons and its 
electromagnetic decays. Sec.~\ref{sec:Strongdecays} is devoted to the study of strong decays and reactions, whereas in 
Sec.~\ref{sec:WeakDecays}, we perform the study of the B weak decays into open charm states. We end by summarizing the work and giving
some conclusions in Sec.~\ref{sec:Summary}. 

%%%%%%%%%%%%%%%%%%%%%%%%%%%%%%%%%%%%%%%%%%%%%%%%%%%%%%%%%%%%%%%%%%%%%%%%%%%%%%%%
\FloatBarrier
%%%%%%%%%%%%%%%%%%%%%%%%%%%%%%%%%%%%%%%%%%%%%%%%%%%%%%%%%%%%%%%%%%%%%%%%%%%%%%%%

\section{Constituent Quark Model}
\label{sec:ConQuaMod}

Constituent quark models have a long history starting from the Isgur seminal work (see, for example 
Refs.~\refcite{PhysRevD.19.2653} and~\refcite{doi:10.1142/S0218301392000242}) 
in which the potential 
between two massive quarks (constituents) was modeled by a quadratic confinement potential plus a chromomagnetic interaction. This model was 
successful in explaining the baryon and meson spectra known at that time. In the eighties it was realized that the constituent mass is a 
consequence of the chiral symmetry breaking in the light quark sector at a momentum scale $\Lambda_{sb}$ greater than the confinement 
scale $\Lambda_{conf}$~\cite{Manohar1984189}. In the region between the two scales, due to this breaking, the quark propagator gets modified 
and quarks acquire a dynamical momentum dependent mass~\cite{Dyakonov1986457}.
The Lagrangian describing this scenario must contains chiral fields to compensate the mass term. 
The Goldstone bosons associated to the chiral fields leads to an additional interaction between light quarks. 
This fact does not affect to the heavy quark sector but is of paramount importance in the molecular picture because 
the only remaining  interaction between the two molecular components, due to its color singlet nature, is the one driven 
by the Goldstone boson exchanges between the light quarks.

The simplest Lagrangian which
contain chiral fields to compensate the mass term can be expressed as 
\begin{equation}\label{lagrangian}
{\mathcal L}
=\overline{\psi }(i\, {\slash\!\!\! \partial} -M(q^{2})U^{\gamma_{5}})\,\psi 
\end{equation}
where $U^{\gamma _{5}}=\exp (i\pi ^{a}\lambda ^{a}\gamma _{5}/f_{\pi })$,
$\pi ^{a}$ denotes nine pseudoscalar fields $(\eta _{0,}\vec{\pi }
,K_{i},\eta _{8})$ with $i=$1,...,4 and $M(q^2)$ is the constituent mass.
This constituent quark mass, which vanishes at large momenta and is frozen at low 
momenta at a value around 300 MeV, can be explicitly obtained from the theory but
its theoretical behavior can be simulated by parametrizing 
$M(q^{2})=m_{q}F(q^{2})$ where $m_{q}\simeq $ 300 MeV, and
\begin{equation}
F(q^{2})=\left[ \frac{{\Lambda}^{2}}{\Lambda ^{2}+q^{2}}
\right] ^{\frac{1}{2}} \, .
\end{equation} 
The cut-off $\Lambda$ fixes the
chiral symmetry breaking scale.

The Goldstone boson field matrix $U^{\gamma _{5}}$ can be expanded in terms of boson fields,
\begin{equation}
U^{\gamma _{5}}=1+\frac{i}{f_{\pi }}\gamma ^{5}\lambda ^{a}\pi ^{a}-\frac{1}{%
2f_{\pi }^{2}}\pi ^{a}\pi ^{a}+...
\end{equation}
The first term of the expansion generates the constituent quark mass, while the
second gives rise to a one-boson exchange interaction between quarks. The
main contribution of the third term comes from the two-pion exchange which
has been simulated by means of a scalar exchange potential.

In the heavy quark sector 
chiral symmetry is explicitly broken and this type of interaction does not act. 
However it constrains the model parameters through the light meson phenomenology 
and provides a natural way to incorporate the pion exchange interaction in the 
molecular dynamics. 

Beyond the chiral symmetry breaking scale one 
expects the dynamics to be governed by QCD perturbative effects.
They are taken 
into account through the one gluon-exchange interaction.

The one-gluon exchange potential is generated from the vertex Lagrangian
\begin{equation}
\label{Lqqg}
{\mathcal L}_{qqg} = i\sqrt{4\pi\alpha_{s}} \bar{\psi} \gamma_{\mu} G^{\mu}_{c}
\lambda^{c} \psi,
\end{equation}
where $\lambda^{c}$ are the $SU(3)$ color matrices and $G^{\mu}_{c}$ is the gluon
field. The resulting potential contains central, tensor and spin-orbit contributions
given by
\begin{equation}
\begin{split}
&
V_{\rm OGE}^{\rm C}(\vec{r}_{ij}) =
\frac{1}{4}\alpha_{s}(\vec{\lambda}_{i}^{c}\cdot
\vec{\lambda}_{j}^{c})\left[ \frac{1}{r_{ij}}-\frac{1}{6m_{i}m_{j}} 
(\vec{\sigma}_{i}\cdot\vec{\sigma}_{j}) 
\frac{e^{-r_{ij}/r_{0}(\mu)}}{r_{ij}r_{0}^{2}(\mu)}\right], \\
& 
V_{\rm OGE}^{\rm T}(\vec{r}_{ij})=-\frac{1}{16}\frac{\alpha_{s}}{m_{i}m_{j}}
(\vec{\lambda}_{i}^{c}\cdot\vec{\lambda}_{j}^{c})\left[ 
\frac{1}{r_{ij}^{3}}-\frac{e^{-r_{ij}/r_{g}(\mu)}}{r_{ij}}\left( 
\frac{1}{r_{ij}^{2}}+\frac{1}{3r_{g}^{2}(\mu)}+\frac{1}{r_{ij}r_{g}(\mu)}\right)
\right]S_{ij}, \\
&
\begin{split}
V_{\rm OGE}^{\rm SO}(\vec{r}_{ij})= &  
-\frac{1}{16}\frac{\alpha_{s}}{m_{i}^{2}m_{j}^{2}}(\vec{\lambda}_{i}^{c} \cdot
\vec{\lambda}_{j}^{c})\left[\frac{1}{r_{ij}^{3}}-\frac{e^{-r_{ij}/r_{g}(\mu)}}
{r_{ij}^{3}} \left(1+\frac{r_{ij}}{r_{g}(\mu)}\right)\right] \times \\ & \times 
\left[((m_{i}+m_{j})^{2}+2m_{i}m_{j})(\vec{S}_{+}\cdot\vec{L})+
(m_{j}^{2}-m_{i}^{2}) (\vec{S}_{-}\cdot\vec{L}) \right],
\end{split}
\end{split}
\end{equation}
where $r_{0}(\mu_{ij})=\hat{r}_{0}\frac{\mu_{nn}}{\mu_{ij}}$ and
$r_{g}(\mu_{ij})=\hat{r}_{g}\frac{\mu_{nn}}{\mu_{ij}}$ are regulators. Note that the
contact term of the central part of the one-gluon exchange potential has been
regularized as follows
\begin{equation}
\delta(\vec{r}_{ij})\sim\frac{1}{4\pi
r_{0}^{2}}\frac{e^{-r_{ij}/r_{0}}}{r_{ij}}.
\label{eq:delta}
\end{equation}

To improve the description of mesons with different flavored quarks we include 
one-loop corrections to the OGE potential
as derived by Gupta {\it et al.}~\cite{PhysRevD.24.2309}. 
\begin{equation}
\begin{split}
&
V_{\rm OGE}^{\rm 1-loop,C}(\vec{r}_{ij})=0, \\
&
\begin{split}
V_{\rm OGE}^{\rm 1-loop,T}(\vec{r}_{ij}) = \frac{C_{F}}{4\pi}
\frac{\alpha_{s}^{2}}{m_{i}m_{j}}\frac{1}{r^{3}}S_{ij} 
&
\left[\frac{b_{0}}{2}\left(\ln(\mu
r_{ij})+\gamma_{E}-\frac{4}{3}\right)+\frac{5}{12}b_ {0}-\frac{2}{3}C_{A}
\right. \\ & \left.
+\frac{1}{2}\left(C_{A}+2C_{F}-2C_{A}\left(\ln(\sqrt{m_{i}m_{j}}\,r_{ij}
)+\gamma_ {
E}-\frac{4}{3}\right)\right)\right],
\end{split} \\
&
\begin{split}
&V_{\rm OGE}^{\rm 1-loop,SO}(\vec{r}_{ij})=\frac{C_{F}}{4\pi}
\frac{\alpha_{s}^{2}}{m_{i}^{2}m_{j}^{2}}\frac{1}{r^{3}}\times \\
&
\begin{split}
\times\Bigg\lbrace (\vec{S}_{+}\cdot\vec{L}) & \Big[
\left((m_{i}+m_{j})^{2}+2m_{i}m_{j}\right)\left(C_{F}+C_{A}-C_{A}
\left(\ln(\sqrt{m_{i}m_{j}}\,r_{ij})+\gamma_{E}\right)\right) \\
&
+4m_{i}m_{j}\left(\frac{b_{0}}{2}\left(\ln(\mu
r_{ij})+\gamma_{E}\right)-\frac{1}{12}b_{0}-\frac{1}{2}C_{F}-\frac{7}{6}C_{A}
+\frac{C_{A}}{2}\left(\ln(\sqrt{m_{i}m_{j}}\,r_{ij})+\gamma_{E}\right)\right)
\\
&
+\frac{1}{2}(m_{j}^{2}-m_{i}^{2})C_{A}\ln\left(\frac{m_{j}}{m_{i}}\right)\Big] 
\end{split} \\
&
\begin{split}
\,\,\,\,\,\,\,+(\vec{S}_{-}\cdot\vec{L}) &
\Big[(m_{j}^{2}-m_{i}^{2})\left(C_{F}+C_{A}-C_{A}\left(\ln(\sqrt{m_{i}m_{j}}\,r_
{ij})+\gamma_{E}\right)\right) \\
&
+\frac{1}{2}(m_{i}+m_{j})^{2}C_{A}\ln\left(\frac{m_{j}}{m_{i}}\right)\Big]
\Bigg\rbrace,
\end{split}
\end{split}
\end{split}
\end{equation}
where $C_{F}=4/3$, $C_{A}=3$, $b_{0}=9$, $\gamma_{E}=0.5772$ and the scale
$\mu\sim1\,{\rm GeV}$.

Although there is no analytical proof, it is a general belief that confinement
emerges from the force between the gluon color charges. When two quarks are
separated, due to the non-Abelian character of the theory, the gluon fields
self-interact forming color strings which bring the quarks together.

In a pure gluon gauge theory the potential energy of the $q\bar{q}$ pair grows 
linearly with the quark-antiquark distance. However, in full QCD the presence of
sea quarks may soften the linear potential, due to the screening of the color
charges, and eventually leads to the breaking of the string. 
 This characteristic can be translated
into a screened potential in such a way that the potential
saturates at the same interquark distance.
\begin{equation}
\begin{split}
&
V_{\rm CON}^{\rm C}(\vec{r}_{ij})=\left[-a_{c}(1-e^{-\mu_{c}r_{ij}})+\Delta
\right] (\vec{\lambda}_{i}^{c}\cdot\vec{\lambda}_{j}^{c}), \\
&
\begin{split}
&
V_{\rm CON}^{\rm SO}(\vec{r}_{ij}) =
-(\vec{\lambda}_{i}^{c}\cdot\vec{\lambda}_{j}^{c}) \frac{a_{c}\mu_{c}e^{-\mu_{c}
r_{ij}}}{4m_{i}^{2}m_{j}^{2}r_{ij}} \times \\
&
\times \left[((m_{i}^{2}+m_{j}^{2})(1-2a_{s})+4m_{i}m_{j}(1-a_{s}))(\vec{S}_{+}
\cdot\vec{L}) \right. \\
&
\left. \quad\,\, +(m_{j}^{2}-m_{i}^{2}) (1-2a_{s}) (\vec{S}_{-}\cdot\vec{L})
\right],
\end{split}
\end{split}
\end{equation}
where $a_{s}$ controls the mixture between the scalar and vector Lorentz structures
of the confinement. At short distances this potential presents a linear behavior with
an effective confinement strength,
$\sigma=-a_{c}\,\mu_{c}\,(\vec{\lambda}^{c}_{i}\cdot \vec{\lambda}^{c}_{j})$, while
it becomes constant at large distances. This type of potential shows a threshold
defined by
\begin{equation}
V_{\rm thr}=\{-a_{c}+ \Delta\}(\vec{\lambda}^{c}_{i}\cdot
\vec{\lambda}^{c}_{j}).
\end{equation}
No $q\bar{q}$ bound states can be found for energies higher than this threshold.
The system suffers a transition from a color string configuration between two
static color sources into a pair of static mesons due to the breaking of the
color string and the most favored decay into hadrons.

Among the different methods to solve the Schr\"odinger equation and
find the quark-antiquark bound states, we use the Gaussian Expansion
Method~\cite{Hiyama:2003cu} because it provides enough accuracy and it makes the
subsequent evaluation of the decay amplitude matrix elements easier. 

This procedure provides the radial wave function solution of the Schr\"odinger
equation as an expansion in terms of basis functions
\begin{equation}
R_{\alpha}(r)=\sum_{n=1}^{n_{max}} c_{n}^\alpha \phi^G_{nl}(r),
\end{equation} 
where $\alpha$ refers to the channel quantum numbers. The coefficients,
$c_{n}^\alpha$, and the eigenvalue, $E$, are determined from the Rayleigh-Ritz
variational principle
\begin{equation}
\sum_{n=1}^{n_{max}} \left[\left(T_{n'n}^\alpha-EN_{n'n}^\alpha\right)
c_{n}^\alpha+\sum_{\alpha'}
\ V_{n'n}^{\alpha\alpha'}c_{n}^{\alpha'}=0\right],
\end{equation}
where $T_{n'n}^\alpha$, $N_{n'n}^\alpha$ and $V_{n'n}^{\alpha\alpha'}$ are the 
matrix elements of the kinetic energy, the normalization and the potential, 
respectively. $T_{n'n}^\alpha$ and $N_{n'n}^\alpha$ are diagonal whereas the
mixing between different channels is given by $V_{n'n}^{\alpha\alpha'}$.

\begin{table}[!t]
\begin{center}
\caption{\label{tab:parameters} Model parameters fitted over all meson spectra and
relevant for the heavy quark sectors.}
\begin{tabular}{ccc}
\toprule
Quark masses    & $m_{n}$ (MeV) & $313$ \\
		& $m_{s}$ (MeV) & $555$ \\
		& $m_{c}$ (MeV) & $1763$ \\
		& $m_{b}$ (MeV) & $5110$ \\
\colrule
OGE & $\alpha_{0}$ & $2.118$ \\
    & $\Lambda_{0}$ $(\mbox{fm}^{-1})$ & $0.113$ \\
    & $\mu_{0}$ (MeV) & $36.976$ \\
    & $\hat{r}_{0}$ (fm) & $0.181$ \\
    & $\hat{r}_{g}$ (fm) & $0.259$ \\
\colrule
Confinement & $a_{c}$ (MeV) & $507.4$ \\
	    & $\mu_{c}$ $(\mbox{fm}^{-1})$ & $0.576$ \\
	    & $\Delta$ (MeV) & $184.432$ \\
	    & $a_{s}$ & $0.81$ \\
\botrule
\end{tabular}
\end{center}
\end{table}

Following Ref.~\refcite{Hiyama:2003cu}, we employ Gaussian trial functions with
ranges in geometric progression. This enables the optimization of ranges employing a
small number of free parameters. Moreover, the geometric progression is dense at
short distances, so that it allows the description of the dynamics mediated by short
range potentials. The fast damping of the gaussian tail is not a problem, since we
can choose the maximal range much longer than the hadronic size.

Table~\ref{tab:parameters} shows the model parameters fitted over all meson spectra
and relevant for the heavy quark sectors, which have been taken from Ref.~\refcite{PhysRevD.78.114033}.

%%%%%%%%%%%%%%%%%%%%%%%%%%%%%%%%%%%%%%%%%%%%%%%%%%%%%%%%%%%%%%%%%%%%%%%%%%%%%%%%
\FloatBarrier
%%%%%%%%%%%%%%%%%%%%%%%%%%%%%%%%%%%%%%%%%%%%%%%%%%%%%%%%%%%%%%%%%%%%%%%%%%%%%%%%

\section{Spectroscopy}
\label{sec:Spectroscopy}
\subsection{Charmonium}

\begin{table}[!t]
\begin{center}
\caption{\label{tab:predmassescc} Masses, in MeV, of charmonium states. Some
tentative $XYZ$ assignments attending to the masses have been done. The experimental
masses are taken from Particle Data Group (PDG)~\protect\cite{PDG2012} for the well
established states and from their respective original works for $XYZ$ mesons.
We compare our results (labeled as The.) with those predicted by other significant quark models in the
literature: S. Godfrey and N. Isgur~\protect\cite{PhysRevD.32.189}; and D. Ebert,
R.N. Faustov and V.O. Galkin~\protect\cite{Ebert:2011jc}.}
\begin{tabular}{cccccccc}
\toprule
Ref. & Assignment & $J^{PC}$ & nL & The. & Ref.~\refcite{PhysRevD.32.189} & Ref.~\refcite{Ebert:2011jc} & Exp. \\
\colrule
\refcite{PDG2012} & $\eta_{c}(1S)$ & $0^{-+}$ & $1S$ & $2990$ & $2970$ & $2981$ & $2981.0\pm1.1$ \\
\refcite{PDG2012} & $\eta_{c}(2S)$ &          & $2S$ & $3643$ & $3620$ & $3635$ & $3638.9\pm1.3$ \\
                  &                &          & $3S$ & $4054$ & $4060$ & $3989$ & - \\[2ex]
\refcite{PDG2012} & $\chi_{c0}(1P)$ & $0^{++}$ & $1P$ & $3452$ & $3440$ & $3413$ & $3414.75\pm0.31$ \\
\refcite{PhysRevLett.104.092001} & $X(3915)$ & & $2P$ & $3909$ & $3920$ & $3870$ & $3915\pm3\pm2$ \\
                  &                          & & $3P$ & $4242$ & -      & $4301$ & - \\[2ex]
\refcite{PDG2012} & $h_{c}(1P)$ & $1^{+-}$ & $1P$ & $3515$ & $3520$ & $3525$ & $3525.41\pm0.16$ \\
                  &             &          & $2P$ & $3956$ & $3960$ & $3926$ & - \\
                  &             &          & $3P$ & $4278$ & -      & $4337$ & - \\[2ex]
\refcite{PDG2012} & $J/\psi$ & $1^{--}$        & $1S$ & $3096$ & $3100$ & $3096$ & $3096.916\pm0.011$ \\
\refcite{PDG2012} & $\psi(2S)$               & & $2S$ & $3703$ & $3680$ & $3685$ & $3686.108\pm0.018$ \\
\refcite{PDG2012} & $\psi(3770)$             & & $1D$ & $3796$ & $3820$ & $3783$ & $3778.1\pm1.2$ \\
\refcite{PDG2012} & $\psi(4040)$             & & $3S$ & $4097$ & $4100$ & $4039$ & $4039\pm1$ \\
\refcite{PDG2012} & $\psi(4160)$             & & $2D$ & $4153$ & $4190$ & $4150$ & $4153\pm3$ \\
\refcite{PhysRevLett.99.142002} & $X(4360)$  & & $4S$ & $4389$ & $4450$ & $4427$ & $4361\pm9\pm9$ \\
\refcite{PDG2012} & $\psi(4415)$             & & $3D$ & $4426$ & $4520$ & $4507$ & $4421\pm4$ \\
\refcite{PhysRevLett.101.172001} & $X(4630)$ & & $5S$ & $4614$ & -      & $4837$ & $4634^{+8+5}_{-7-8}$ \\
\refcite{PhysRevLett.99.142002} & $X(4660)$  & & $4D$ & $4641$ & -      & $4857$ & $4664\pm11\pm5$ \\[2ex]
\refcite{PDG2012} & $\chi_{c1}(1P)$ & $1^{++}$ & $1P$ & $3504$ & $3510$ & $3511$ & $3510.66\pm0.07$ \\
                  &                 &          & $2P$ & $3947$ & $3950$ & $3906$ & - \\
                  &                 &          & $3P$ & $4272$ & -      & $4319$ & - \\[2ex]
& $\eta_{c2}(1D)$ & $2^{-+}$ & $1D$ & $3812$ & $3840$ & $3807$ & - \\
&                 &          & $2D$ & $4166$ & $4210$ & $4196$ & - \\
&                 &          & $3D$ & $4437$ & -      & $4549$ & - \\[2ex]
\refcite{PDG2012} & $\chi_{c2}(1P)$ & $2^{++}$ & $1P$ & $3532$ & $3550$ & $3555$ & $3556.20\pm0.09$ \\
\refcite{PhysRevLett.96.082003} & $Z(3930)$ &  & $2P$ & $3969$ & $3980$ & $3949$ & $3929\pm5\pm2$  \\
                                &           &  & $1F$ & $4043$ & $4010$ & $4041$ & - \\[2ex]
\refcite{Bhardwaj:2013rmw} & $X(3823)$ & $2^{--}$ & $1D$ & $3810$ & $3840$ & $3795$ & $3823.1\pm1.8\pm0.7$ \\
                           &           &          & $2D$ & $4164$ & $4210$ & $4190$ & - \\
                           &           &          & $3D$ & $4436$ & -      & $4544$ & - \\
\botrule
\end{tabular}
\end{center}
\end{table}

Shortly after the discovery by BELLE of the missing $\eta'_{c}(2^1S_0)$~\cite{PhysRevLett.89.102001},
new states containing charm quarks have appeared in great profusion. Some of them
have been identified as canonical $c\bar c$ states. but others, called collectively as XYZ states, 
exhibit unexpected properties which hardly fit with those of two quark states. 

The charmonium spectrum is given in Table~\ref{tab:predmassescc}. We compare our
results with the experimental data and with those predicted by other significant
quark models in the literature: S. Godfrey and N. Isgur~\cite{PhysRevD.32.189}; 
and D. Ebert, R.N. Faustov and V.O. Galkin~\cite{Ebert:2011jc}. Some tentative 
$XYZ$ assignments attending to the masses have been done. The experimental masses 
are taken from Particle Data Group (PDG)~\cite{PDG2012} for the well established 
states and from their respective original works for $XYZ$ mesons.

As one can see in Table~\ref{tab:predmassescc}, we obtain a quite reasonable global
description of the charmonium sector. This feature is also reached by other quark
models. The spectrum predicted by the different models is quite similar at least for the low lying levels.
However, while our confining term is based on a screened linear
potential at large interquark distances, the remaining models implement a linear
potential for all distances. This can be translated into a different prediction of
the masses for the higher excited states, the screened linear potential reduces the
masses of higher excited states, see Table~\ref{tab:predmassescc}. This has
an important consequence, the new assignment of the $\psi(4415)$. Usually this state
has been assigned as a $4S$ state. Our particular choice of the potential includes
the new $X(4360)$ as a $4S$ state between the well established $\psi(4160)$ and
$\psi(4415)$ which are both predicted as $D$-wave states. Moreover we can assign as 
$1^{--}$ $c\bar{c}$ structures the new $X(4630)$ and $X(4660)$ mesons whose
nature is still unclear.

It is important to remark that a nonrelativistic treatment of the quark-antiquark
system is performed in our approach. However, a relativistic scheme is used for the
quark models of Refs.~\refcite{PhysRevD.32.189} and~\refcite{Ebert:2011jc}. The
relativistic effects should be small due to the large mass of the $c$-quark. Therefore,
the differences on the spectrum between both schemes are negligible and can be
absorbed in the reparametrization of the model. 

The $\eta_{c}(1S)$ is the lowest state of charmonium. The model predicts a mass of
$2990\,{\rm MeV}$, in good agreement with the experimental one. The splitting between
$1^{1}S_{0}$ and $1^{3}S_{1}$ is given by the Dirac delta term of the OGE potential.
This splitting is measured experimentally to be $116.6\pm1.2\,{\rm MeV}$ which is in
reasonable agreement with our prediction of $106\,{\rm MeV}$. Moreover, our predicted mass for the
$\eta_{c}(2S)$ is $3643\,{\rm MeV}$, which agrees with the experimental value.

Lattice data show a vanishing long-range component of the spin-spin potential. Thus,
this part of the potential appears to be entirely dominated by its short-range,
delta-like term, suggesting that the $^{1}P_{1}$ should be close to the
center-of-gravity of the $^{3}P_{J}$ system. The precision measurement of
the $h_{c}(1P)$ mass was reported by CLEO in $2008$~\cite{PhysRevLett.101.182003},
$3525.28\pm0.19\pm0.12\,{\rm MeV}$. Later, BES~III~\cite{PhysRevLett.104.132002} has
confirmed this with a mass of $3525.40\pm0.13\pm0.18\,{\rm MeV}$. The centroid of the
$1^3P_J$ states is known to be $3525.30\pm0.04\,{\rm MeV}$~\cite{PDG2012} and then
the hyperfine splitting is $+0.02\pm0.23\,{\rm MeV}$ from CLEO and
$-0.10\pm0.22\,{\rm MeV}$ from BES~III. The comparison in our model between the
centroid of $^{3}P_{J}$ states and the corresponding $h_{c}$ mass shows that our
spin-spin interaction is negligible for these channels and it is in perfect agreement
with the lattice expectations and the experimental measurements for the ground state.

As shown in Table~\ref{tab:predmassescc} the long known $1^{3}P_{J}$ states are in
agreement with the model results. The mean $2P$ multiplet mass is predicted to be
near $3.95\,{\rm GeV}$. Although no $2P$ $c\bar{c}$ state has been clearly seen
experimentally, there are reports from the different Collaborations which claim
enhancements in that energy region. Among them one can cite the $X(3872)$, $X(3915)$, $Y(3940)$,
$X(3940)$ and $Z(3930)$.

The $X(3872)$ mass is difficult to reproduce by the standard quark models, see
Table~\ref{tab:predmassescc}. The $X(3872)$ mass is extremely close to the
$D^{0}D^{\ast0}$ threshold so it appears as a natural candidate to an even $C$-parity
$D^{0}D^{\ast0}$ molecule. The molecular interpretation will also explain the large
isospin violation, but runs into trouble when it tries to explain the high
$\gamma\psi'$ decay rate. This puzzling situation suggests for the $X(3872)$ state a
combination of a $2P$ $c\bar{c}$ state and a weakly-bound $D^{0}D^{\ast0}$
molecule. In Ref.~\refcite{PhysRevD.81.054023} we have performed a coupled channel
calculation of the $1^{++}$ $c\bar{c}$ sector including $q\bar{q}$ and
$q\bar{q}q\bar{q}$ configurations. Two and four quark configurations are coupled
nonperturbatively using the $^{3}P_{0}$ model. The elusive $X(3872)$ meson appears as
a new state with a high probability for the $DD^{\ast}$ molecular component. The
original $c\bar{c}(2^{3}P_{1})$ state acquires a sizable $DD^{\ast}$ component and
can be identified with the $X(3940)$.

The $Y(3940)\rightarrow\omega J/\psi$ enhancement was initially found by 
Belle~\cite{PhysRevLett.94.182002} in $B^{+}\rightarrow K^{+}Y(3940)$ decays. It
was confirmed by BaBar~\cite{PhysRevLett.101.082001} with more statistics,
albeit with somewhat smaller mass. But Belle~\cite{PhysRevLett.104.092001} also
found a statistically compelling resonant structure $X(3915)$ in $\gamma\gamma$
fusion decaying to $\omega J/\psi$. It shares the same production and decay
signature as that of BaBar's $Y(3940)$, which has mass and width consistent
with the $X(3915)$. An interpretation of these two states as been the same appears
as a widely accepted idea and the name which is conserved is $X(3915)$. We only
know at the moment that this state has an even $C$-parity. If $X(3915)$ was a
$c\bar{c}$ state, the most probable quantum numbers would be $0^{++}$. The
mass predicted for the $2^{3}P_{0}$ is $3909$, in very good agreement with the
experimental measurement.

In 2005 Belle~\cite{PhysRevLett.96.082003} observed an enhancement in the
$D\bar{D}$ mass spectrum from $e^{+}e^{-}\rightarrow e^{+}e^{-}D\bar{D}$ events
with a statistical significance of $5.3\sigma$. It was initially dubbed the
$Z(3930)$, but since then it has been widely (if not universally) accepted as
the $\chi_{c2}(2P)$. There is some Lattice calculations~\cite{PhysRevD.79.094504}
which suggest that the $\chi_{c2}(2P)$ and the $1^{3}F_{2}$ state could be quite
close in mass, so that perhaps the $Z(3930)$ is not the $2^{3}P_{2}$ but rather the
$1^{3}F_{2}$. Table~\ref{tab:predmassescc} shows that all quark models predict a
mass splitting between both states of about tens of MeV, so we do not consider that
those states are nearby degenerated and assign the $Z(3930)$ as the $2^{3}P_{2}$
state.

The Belle Collaboration has recently reported measurements of $B\to \chi_{c1}\gamma
K$ and $\chi_{c2}\gamma K$~\cite{Bhardwaj:2013rmw}. They found evidence of a new
resonance in the $\chi_{c1}\gamma$ final state with a mass of
$(3823.1\pm1.8\pm0.7)\,{\rm MeV}$, a value which is consistent with the $1^{3}D_{2}$
$c\bar{c}$ state according to our model, $3812\,{\rm MeV}$. We expect that the
$1^{1}D_{2}$ state appears in the same energy range of the $X(3823)$, however, as we
will see below, this state should appear in the $h_{c}\gamma$ channel.

%%%%%%%%%%%%%%%%%%%%%%%%%%%%%%%%%%%%%%%%%%%%%%%%%%%%%%%%%%%%%%%%%%%%%%%%%%%%%%%%
\FloatBarrier
%%%%%%%%%%%%%%%%%%%%%%%%%%%%%%%%%%%%%%%%%%%%%%%%%%%%%%%%%%%%%%%%%%%%%%%%%%%%%%%%
     
\subsection{Charmed and charmed-strange mesons}

A simple analysis about the properties of hadrons containing a single heavy quark
$Q=c,\,b$ can be carried out in the $m_{Q}\to\infty$ limit. In such a limit, the
heavy quark acts as a static color source for the rest of the hadron, its spin
$\vec{s}_{Q}$ is decoupled from the total angular momentum of the light degrees of
freedom $\vec{j}_{q}=\vec{s}_{q}+\vec{l}$, and they are separately conserved. Heavy
mesons can be organized in doublets, each one corresponding to a particular value of
$j_{q}$ and parity. The lowest lying $Q\bar{q}$ mesons correspond to $l=0$ ($S$-wave
states of the quark model) with $j_{q}^{P}=\frac{1}{2}^{-}$. This doublet comprises
two states with spin-parity $J^{P}=(0^{-},1^{-})$. For $l=1$ ($P$-wave states of the
quark model), it could be either $j_{q}^{P}=\frac{1}{2}^{+}$ or
$j_{q}^{P}=\frac{3}{2}^{+}$, the two corresponding doublets having
$J^{P}=(0^{+},1^{+})$ and $J^{P}=(1^{+},2^{+})$.

However the experimental results show intriguing aspects which contradict this analysis, specially in the charm strange sector.
The abnormally light mass of the mesons $D^*_{s0}(2317)$ and $D_{s1}(2460)$ below the $DK$ and $D^*K$ thresholds respectively make these states very narrow since the only allowed decays violate isospin. The unexpected feature of these mesons is that they have masses close (or even lower) than their charmed partners. Moreover the masses predicted  by most of the theoretical approaches are considerably heavier than the experimental ones

Very recently new $D$ and $D_s$ resonances has been discovered. Thus BaBar collaboration~\cite{PhysRevD.82.111101} 
reported four new resonances: $D(2550)^0$, $D(2600)^0$, $D(2750)^0$ and $D^*(2760)^0$. These resonances has been recently confirmed by 
LHCb collaboration~\cite{Aaij:2013sza} adding two more states $D(3000)^0$ and $D^*(3000)^+$. The results of both collaboration are 
compatibles except in the case of the width of the $D(2600)^0$ measured as $\Gamma=93\pm 6\pm 13)$ by BaBar collaboration 
and $\Gamma=140\pm 17\pm 18)$ by LHCb. Concerning the charmed strange sector three new states has been reported~\cite{PhysRevD.80.092003}: 
$D^*_{s1}(2710)^+$, $D^*_{sJ}(2860)^+$ and $D^*_{sJ}(3040)^+$. These states has been also confirmed by LHCb collaboration~\cite{Aaij:2012pc}.

The spectra of $D$ and $D_s$ are given in Table~\ref{tab:1loopDmesons} and Table~\ref{tab:1loopDsmesons}. We compare our
results   with those predicted by other significant
quark models in the literature: 
D. Ebert, R.N. Faustov and V.O. Galkin~\cite{Ebert:2009ua} and Di Pierro and Eichten~\cite{PhysRevD.64.114004}. 
Assignments for the well established states taken from Particle Data Group (PDG)~\cite{PDG2012} are also given.

The masses predicted for the $0^{-}$ and $1^{-}$ states -- the
$j_{q}^{P}=\frac{1}{2}^{-}$ doublet -- agree with the experimental
measurements in both sectors. The doublet $j_{q}^{P}=\frac{3}{2}^{+}$, which
corresponds to the $2^{+}$ state and one of the low lying $1^{+}$ states, is in
reasonable agreement with experiment.

As one can see, most of the models cannot
reproduce the mass splittings between the $D_{s0}^{\ast}(2317)$, $D_{s1}(2460)$ and
$D_{s1}(2536)$ mesons. This feature is shared by other quark models,  but also by other approaches like lattice QCD
calculations~\cite{PhysRevD.68.071501}.
The charmed and charmed-strange $0^{+}$ states
are sensitive to the one-loop corrections of the OGE potential included in our model which bring their
masses closer to experiment. This is in agreement with the conclusion of Ref.~\refcite{Lakhina2007159}. 
However, their contribution are not enough to solve the puzzle in the
$1^{+}$ sector. The importance of the meson-meson continuum in the $1^{+}$ $c\bar{s}$
sector will be studied later.

\begin{table}[!t]
\begin{center}
\caption{\label{tab:1loopDmesons} Masses, in MeV, of charmed 
mesons predicted by the constituent quark model.
We compare our results with those of other significant quark models in the literature
from Refs.~\protect\refcite{Ebert:2009ua} and  \protect\refcite{PhysRevD.64.114004}.
The experimental data are from the PDG~\cite{PDG2012}.}
\begin{tabular}{cccccc}
\toprule
 Assignment & $J^{P}$ & The. & Ref.~\refcite{Ebert:2009ua} & Ref.~\refcite{PhysRevD.64.114004} & Exp. \\
\colrule
  $D$           & $0^{-}$ & $1896$ & $1871$ & $1868$ & $1867.7\pm0.3$ \\
  $D(2550)$     &         & $2695$ & $2581$ & $2589$ & $2539.4\pm4.5\pm6.8$ \\
                &         & $3154$ & $3062$ & $3141$ &                \\
  $D^*$         & $1^{-}$ & $2014$ & $2010$ & $2005$ & $2010.25\pm0.14$ \\
                &         & $2754$ & $2632$ & $2692$ &                  \\
                &         & $2905$ & $3096$ & $3226$ &                  \\
  $D_0(2400)$   & $0^{+}$ & $2362$ & $2406$ & $2377$ & $2318\pm 29$ \\
                &         & $2925$ & $2919$ & $2949$ &              \\
                &         & $3292$ &        &        &              \\
  $D_1(2420)$   & $1^{+}$ & $2499$ & $2426$ & $2417$ & $2421.4 \pm 0.6$ \\
  $D_1(2430)$   &         & $2535$ & $2469$ & $2426$ & $2427\pm 26\pm 25$ \\
                &         & $3033$ & $2932$ & $2995$ &              \\
  $D_2(2460)$   & $2^{+}$ & $2544$ & $2460$ & $2460$ & $2462.6\pm0.6$ \\
                &         & $3059$ & $3012$ & $3035$ &                \\
                & $2^{-}$ & $2822$ & $2806$ & $2775$ &                \\
                &         & $2962$ & $2850$ & $2873$ &                \\
                & $3^{+}$ & $3094$ & $2863$ & $2799$ &                \\
                &         & $3240$ & $3335$ &        &                \\
                & $3^{-}$ & $2863$ & $3129$ & $3123$ &                \\
                &         & $3260$ & $3145$ &        &                \\
\botrule
\end{tabular}
\end{center}
\end{table}

\begin{table}[!t]
\begin{center}
\caption{\label{tab:1loopDsmesons} Masses, in MeV, of charmed-strange
mesons predicted by the constituent quark model.
We compare our results with those of other significant quark models in the literature
from Refs.~\protect\refcite{Ebert:2009ua} and  {PhysRevD.64.114004 (2001)}.
The experimental data are from the PDG~\cite{PDG2012}.}
\begin{tabular}{cccccc}
\toprule
 Assignment & $J^{P}$ & The. & Ref.~\refcite{Ebert:2009ua} & Ref.~\refcite{PhysRevD.64.114004} & Exp. \\
\colrule
  $D_s$             & $0^{-}$ & $1984$ & $1969$ & $1965$ & $1968.5\pm0.32$ \\
                    &         & $2729$ & $2688$ & $2750$ &                      \\
                    &         & $3178$ &        & $3259$ &                \\
  $D_s^*$           & $1^{-}$ & $2104$ & $2111$ & $2113$ & $2112.3\pm0.5$ \\
  $D_{s_1}^*(2700)$ &         & $2794$ & $2731$ & $2806$ & $2709.0\pm0.4$   \\
                    &         & $2890$ & $2913$ & $2913$ &                  \\
  $D_{s_0}^*(2317)$ & $0^{+}$ & $2383$ & $2509$ & $2487$ & $2317.8\pm 0.6$ \\
                    &         & $2934$ &        & $3067$ &              \\
                    &         & $3310$ &        &        &              \\
  $D_{s_1}(2460)$   & $1^{+}$ & $2560$ & $2536$ & $2535$ & $2459.6 \pm 0.6$ \\
  $D_{s_1}(2526)$   &         & $2570$ & $2574$ & $2605$ & $2535.12\pm 0.13$ \\
                    &         & $3061$ &        & $3114$ &              \\
  $D_{s_2}(2573)$   & $2^{+}$ & $2609$ & $2571$ & $2581$ & $2571.9\pm0.8$ \\
                    &         & $3094$ & $3142$ & $3157$ &                \\
                    & $2^{-}$ & $2888$ & $2931$ & $2900$ &                \\
                    &         & $2943$ & $2961$ & $2953$ &                \\
                    & $3^{+}$ & $3151$ & $3254$ & $3203$ &                \\
                    &         & $3215$ & $3266$ & $3247$ &                \\
                    & $3^{-}$ & $2922$ & $2971$ & $2925$ &                \\
                    &         & $3304$ & $3469$ &        &                \\
\botrule
\end{tabular}
\end{center}
\end{table}
%%%%%%%%%%%%%%%%%%%%%%%%%%%%%%%%%%%%%%%%%%%%%%%%%%%%%%%%%%%%%%%%%%%%%%%%%%%%%%%%
\FloatBarrier
%%%%%%%%%%%%%%%%%%%%%%%%%%%%%%%%%%%%%%%%%%%%%%%%%%%%%%%%%%%%%%%%%%%%%%%%%%%%%%%%

We postpone the assignment of the new states to the strong decay section were the strong width of these meson will be present in detail

\subsection{Electromagnetic decays}

The knowledge of the leptonic decay width of higher $1^{--}$ charmonium states is
important for several reasons. First of all it allows to test the wave function at
very short distances. Moreover it can help to distinguish between conventional
$c\bar{c}$ mesons and multiquark structures which have much smaller dielectron
widths~\cite{Badalian:1985es}. The leptonic widths are compared in
Table~\ref{tab:lepwidthcc}, we include the recent data reported by the BES
Collaboration in Ref.~\refcite{Ablikim2008315}.

\begin{table}[!t]
\begin{center}
\caption{\label{tab:lepwidthcc} Leptonic decay widths, in keV, of $\psi$ states.}
\begin{tabular}{cccccc}
\toprule
$(nL)$ & State & $M_{\rm The.}$ (MeV) & $\Gamma_{\rm The.}$ (keV) & $\Gamma_{\rm
Exp.}$ (keV) & \\
\colrule
$(1S)$ & $J/\psi$ & $3096$ & $3.93$ & $5.55\pm0.14\pm0.02$ & \cite{PDG2012} \\
$(2S)$ & $\psi(2S)$ & $3703$ & $1.78$ & $2.33\pm0.07$ & \cite{PDG2012} \\
$(1D)$ & $\psi(3770)$ & $3796$ & $0.22$ & $0.22\pm0.05$ &
\cite{Ablikim2008315} \\
$(3S)$ & $\psi(4040)$ & $4097$ & $1.11$ & $0.83\pm0.20$ &
\cite{Ablikim2008315} \\
$(2D)$ & $\psi(4160)$ & $4153$ & $0.30$ &  $0.48\pm0.22$ &
\cite{Ablikim2008315} \\
$(4S)$ & $X(4360)$ & $4389$ & $0.78$ & - & - \\
$(3D)$ & $\psi(4415)$ & $4426$ & $0.33$ & $0.35\pm0.12$ &
\cite{Ablikim2008315} \\
$(5S)$ & $X(4630)$ & $4614$ & $0.57$ & - & - \\ 
$(4D)$ & $X(4660)$ & $4641$ & $0.31$ & - & - \\
\botrule
\end{tabular}
\end{center}
\end{table}

As we have mentioned, one striking feature of our model is the new assignment
of the $\psi(4415)$. Usually this state has been assigned as a $4S$ state. Our
particular choice of the potential includes the new $X(4360)$ as a $4S$ state between
the well established $\psi(4160)$ and $\psi(4415)$ which are both predicted as
$D$-wave states. Whether or not this assignment is correct can be tested with the
$e^{+}e^{-}$ leptonic widths. From Table~\ref{tab:lepwidthcc} one can see that the
width of the $4S$ state is $0.78\,{\rm keV}$, whereas the experimental value for the
$\psi(4415)$ is $\Gamma_{e^{+}e^{-}}=0.35\pm0.12\,{\rm keV}$, in excellent agreement
with the result for the $3D$ state ($0.33\,{\rm keV}$). The measurement of the
leptonic width for the $X(4360)$ is very important and would clarify the situation.

It is generally assumed that the $1^{--}$ $c\bar{c}$ mesons are a mixture of
$^{3}S_{1}$ and $^{3}D_{1}$ states in order to reproduce the leptonic widths. In our
model the mixing is not fitted to the experimental data but driven by the tensor
piece of the quark-antiquark interaction. All are almost pure states either
$^{3}S_{1}$ or $^{3}D_{1}$ and we can reasonably reproduce the leptonic widths.

The study of higher multipole contributions to the radiative transitions between
spin-triplet states involves an alternative way to disentangle the mixing
between $S$- and $D$-waves in $1^{--}$ $c\bar{c}$ mesons. The radiative decay
sequences
\begin{equation}
e^{+}e^{-} \to \psi(2S), \quad \psi(2S) \to \gamma'\,\chi_{(c1,c2)}, \quad
\chi_{(c1,c2)} \to \gamma J/\psi, \quad J/\psi \to e^{+}e^{-} \mbox{ or }
\mu^{+}\mu^{-},
\end{equation}
has been studied experimentally in Ref.~\refcite{PhysRevD.80.112003}. The electric
dipole amplitudes are dominant but higher multipole contributions are allowed.

\begin{figure}[!t]
\begin{center}
\parbox[c]{0.50\textwidth}{
\centering
\includegraphics[width=0.50\textwidth]{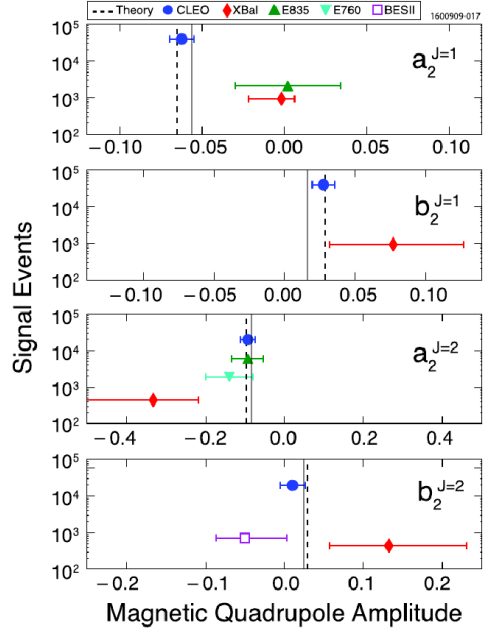}
}
\caption{\label{fig:multipole} Figure from Ref.~\protect\refcite{PhysRevD.80.112003}.
Experimental values of the magnetic quadrupole amplitudes obtained by the CLEO
Collaboration and their comparison with previous experimental data and theoretical
expectations.}
\end{center}
\end{figure}

\begin{table}[!t]
\begin{center}
\caption{\label{tab:sequences} Branching fraction for the decay
$\psi(2S)\rightarrow \gamma(\gamma J/\psi)_{\chi_{cJ}}$. Experimental data are
from Ref.~\protect\refcite{PhysRevD.78.011102}.}
\begin{tabular}{ccc}
\toprule
Mode & $\Gamma_{\rm The.}$ & $\Gamma_{\rm Exp.}$ \\
\colrule
$\gamma(\gamma J/\psi)_{\chi_{c0}}$ & $0.156$ & $0.125\pm0.007\pm0.013$ \\
$\gamma(\gamma J/\psi)_{\chi_{c1}}$ & $4.423$ & $3.56\pm0.03\pm0.12$ \\
$\gamma(\gamma J/\psi)_{\chi_{c2}}$ & $2.099$ & $1.95\pm0.02\pm0.07$ \\
\botrule
\end{tabular}
\end{center}
\end{table}

For the $\chi_{cJ}$ $(J=1,\,2)$ sequences, they search for two multipole amplitudes
$b_{2}^{J=1,\,2}$ and $a_{2}^{J=1,\,2}$, where $b$ stands for the amplitude where
$\chi_{cJ}$ is a reaction product $(\psi'\to\gamma'\chi_{cJ})$ and $a$ stands for the
amplitude where $\chi_{cJ}$ is the decay particle $(\chi_{cJ}\to \gamma J/\psi)$.

We show in Fig.~\ref{fig:multipole} the experimental data (solid circles)
obtained by the CLEO Collaboration in Ref.~\refcite{PhysRevD.80.112003}. The rest
of the data are previous to Ref.~\refcite{PhysRevD.80.112003}. Our theoretical
estimations are represented by a vertical solid line. The same theoretical
estimations considering a $c$-quark mass $(m_{c}=1.5\,{\rm GeV})$ closer to the PDG
value are represented by a vertical dashed line as given in
Ref.~\refcite{PhysRevD.80.112003}. The last experimental measurements and the
theoretical estimations agree well. In some sense it indicates that the
mixing between $S$ and $D$-waves in the $1^{--}$ $c\bar{c}$ states is small, but also
in others as the $2^{++}$ channel where the mixing is between the $P$ and $F$-waves.

To end the above discussion, one can calculate the branching fraction of the process
$\psi(2S)\rightarrow J/\psi \gamma \gamma$ trough $\gamma \chi_{cJ}$. In
Table~\ref{tab:sequences} we compare our results with the most recent
experimental data~\cite{PhysRevD.78.011102}. We reproduce not only the tendency of
the experimental data but also the absolute value.

Table~\ref{tab:radiativeetac2psi2}
shows the E1 radiative decay widths for the first two states of $\eta_{c2}$ and
$\psi_{2}$ that may be useful for experimentalists.
The recently reported~\cite{Bhardwaj:2013rmw} $X(3823)$ state 
has been assigned to the $1^{3}D_{2}$ $c\bar{c}$
state~\ref{tab:predmassescc}. An upper limit of the branching ratio 
${\cal B}(X(3823)\to \chi_{c2}(1P)\gamma)/{\cal B}(\chi_{c1}(1P)\gamma)<0.41$ has been also
given by experimentalists. Our value, $0.24$, is below that limit and assures our
assignment. The reason why the $1^{3}D_{2}$ $c\bar{c}$ state has been difficult to
observe is that open-flavor decay modes are not allowed. The same situation appears
for the $1^{1}D_{2}$ state being the E1 radiative decays the most plausible decay
channels in which this particle can be observed. 

\begin{table}[t!]
\begin{center}
\caption{\label{tab:radiativeetac2psi2} E1 radiative transitions for the first two
states of $\eta_{c2}$ and $\psi_{2}$.}
\begin{tabular}{ccr}
\toprule
Initial meson & Final meson & $\Gamma_{\rm CQM}$ (keV) \\
\colrule
$\eta_{c2}(1^{1}D_{2})$ & $h_{c}(1^{1}P_{1})$ & $276.95$ \\[2ex]
$\eta_{c2}(2^{1}D_{2})$ & $h_{c}(1^{1}P_{1})$ & $114.66$ \\
& $h_{c}(2^{1}P_{1})$ & $211.78$ \\[2ex]
$\psi_{2}(1^{3}D_{2})$ & $\chi_{c1}(1^{3}P_{1})$ & $224.10$ \\
& $\chi_{c2}(1^{3}P_{2})$ & $53.74$ \\[2ex]
$\psi_{2}(2^{3}D_{2})$ & $\chi_{c1}(1^{3}P_{1})$ & $95.44$ \\
& $\chi_{c2}(1^{3}P_{2})$ & $19.92$ \\
& $\chi_{c1}(2^{3}P_{1})$ & $164.35$ \\
& $\chi_{c2}(2^{3}P_{2})$ & $47.92$ \\
& $\chi_{c2}(1^{3}F_{2})$ & $3.88$ \\[2ex]
\botrule
\end{tabular}
\end{center}
\end{table}

%%%%%%%%%%%%%%%%%%%%%%%%%%%%%%%%%%%%%%%%%%%%%%%%%%%%%%%%%%%%%%%%%%%%%%%%%%%%%%%%
\FloatBarrier
%%%%%%%%%%%%%%%%%%%%%%%%%%%%%%%%%%%%%%%%%%%%%%%%%%%%%%%%%%%%%%%%%%%%%%%%%%%%%%%%

\section{Strong Decays}
\label{sec:Strongdecays}

Meson strong decay is a complex nonperturbative process that has not yet been
described from first principles of QCD. This leads to a rather poorly understood area of
hadronic physics which is a problem because decay widths comprise a large portion of
our knowledge of the strong interaction. 

Several phenomenological models have been developed to deal with this topic. The
most popular is the $^{3}P_{0}$ model which assumes that a quark-antiquark pair is
created with vacuum quantum numbers, $J^{PC}=0^{++}$. The $^{3}P_{0}$ model was first
proposed by Micu~\cite{Micu:1968mk}. Le Yaouanc {\it et al.} applied subsequently
this model to meson~\cite{PhysRevD.8.2223} and baryon~\cite{PhysRevD.9.1415}
open-flavor strong decays in a series of publications in the 1970s.

We calculate in this Section the total decay widths of the mesons which belong to
charmed and charmed-strange through a modified version of the
$^{3}P_{0}$ model with a scale dependent strength $\gamma$ of the decay interaction given by .
\begin{equation}
\gamma(\mu) = \frac{\gamma_{0}}{\log\left(\frac{\mu}{\mu_{\gamma}}\right)},
\label{eq:fitgamma}
\end{equation}
where $\mu$ is the reduced mass of the $q\bar{q}$ pair of the decaying meson and
$\gamma_{0}=0.81\pm0.02$ and $\mu_{\gamma}=49.84\pm2.58\,{\rm MeV}$ are 
parameters determined through the fit to some selected the total decay widths. Reference~\refcite{Segovia:2012cd} provides a detailed
explanation on how the fit was performed and on the convention for the definition of
$\gamma$ (see Eq.~(2) of Ref.~\refcite{Segovia:2012cd}).

\begin{table}[!t]
\begin{center}
\caption{\label{tab:totalwidths} Calculated through the $^{3}P_{0}$ model, the strong
total decay widths of the mesons which belong to charmed, charmed-strange, hidden
charm and hidden bottom sectors. The value of the parameter $\gamma$ in every quark
sector is given by Eq.~(\ref{eq:fitgamma}).}
\scalebox{0.85}{\begin{tabular}{ccccccc}
\toprule
Meson &  J & P & C &  Mass (MeV) & $\Gamma_{\rm Exp.}$
(MeV)~\cite{PDG2012} & $\Gamma_{\rm The.}$ (MeV) \\
\colrule
$\psi(3770)$ & $1$ & $-1$ & $-1$ &  $3775.2\pm1.7$ & $27.6\pm1.0$ & $26.5\pm1.7$ \\
$\psi(4040)$ &  $1$ & $-1$ & $-1$ &  $4039\pm1$     & $80\pm10$    & $111.2\pm7.0$ \\
$\psi(4160)$ &  $1$ & $-1$ & $-1$ &  $4153\pm3$     & $103\pm8$    & $115.9\pm7.3$ \\
$X(4360)$    &  $1$ & $-1$ & $-1$ &  $4361\pm9$     & $74\pm18$    & $113.9\pm7.2$ \\
$\psi(4415)$ &  $1$ & $-1$ & $-1$ &  $4421\pm4$     & $119\pm16$~\cite{PhysRevD.72.017501} & $159.0\pm10.0$ \\
$X(4640)$    &  $1$ & $-1$ & $-1$ &  $4634\pm8$     & $92\pm52$    & $206.3\pm13.0$ \\
$X(4660)$    &   $1$ & $-1$ & $-1$ &  $4664\pm11$    & $48\pm15$    & $135.0\pm8.6$ \\
\colrule
$D^{\ast}(2010)^{\pm}$     &  $1$ & $-1$ & - &   $2010.25\pm0.14$ & $0.096\pm0.022$ & $0.036\pm0.003$ \\
$D_{0}^{\ast}(2400)^{\pm}$ &  $0$ & $+1$ & - &   $2403\pm38$      & $283\pm42$      & $212.01\pm17.1$ \\
$D_{1}(2420)^{\pm}$        &  $1$ & $+1$ & - &  $2423.4\pm3.1$   & $25\pm6$        & $25.3\pm2.0$ \\
$D_{1}(2430)^{0}$          &  $1$ & $+1$ & - &  $2427\pm36$      & $384\pm150$     & $229.2\pm18.5$ \\
$D_{2}^{\ast}(2460)^{\pm}$ &  $2$ & $+1$ & - &  $2460.1\pm4.4$   & $37\pm6$        & $64.1\pm5.2$ \\
$D(2550)^{0}$              &  $0$ & $-1$ & - &  $2539.4\pm8.2$   & $130\pm18$      & $132.1\pm10.7$ \\
$D^{\ast}(2600)^{0}$       &  $1$ & $-1$ & - &  $2608.7\pm3.5$   & $93\pm14$       & $96.9\pm7.8$ \\
$D_{J}(2750)^{0}$          &  $2$ & $-1$ & - &  $2752.4\pm3.2$   & $71\pm13$       & $229.9\pm18.6$ \\
$D_{J}^{\ast}(2760)^{0}$   &  $3$ & $-1$ & - &  $2763.3\pm3.3$   & $60.9\pm6.2$    & $116.4\pm9.3$ \\
\colrule
$D_{s1}(2536)^{\pm}$        &  $1$ & $+1$ & - &  $2535.12\pm0.25$ & $1.03\pm0.13$~\cite{aubert2006precision} & $0.99\pm0.07$ \\
$D_{s2}^{\ast}(2575)^{\pm}$ &  $2$ & $+1$ & - &  $2572.6\pm0.9$ & $20\pm5$ & $18.7\pm1.3$ \\
$D_{s1}^{\ast}(2710)^{\pm}$ &  $1$ & $-1$ & - &  $2710\pm14$ & $149\pm65$ & $170.8\pm12.1$ \\
$D_{sJ}^{\ast}(2860)^{\pm}$ &  $\left[\begin{matrix} 1 \\ 3 \end{matrix}\right]$ & $-1$ & - &  $2862\pm6$ & $48\pm7$ & $\left[\begin{matrix} 153.2\pm10.9 \\ 85.1\pm6.1 \end{matrix}\right]$ \\
$D_{sJ}(3040)^{\pm}$        & $1$ & $+1$ & -  & $3044\pm31$ & $239\pm71$ & $ 301.5\pm21.5 $ \\[2ex]
\botrule\end{tabular}}
\end{center}
\end{table}

Table~\ref{tab:totalwidths} shows our results for the total strong decay widths of
the mesons which belong to hidden charm, charmed and charmed-strange sectors. 
In the case of mesons containing a single $c$-quark, we have considered the
newly observed charmed mesons $D(2550)$, $D^{\ast}(2600)$, $D_{J}(2750)$ and
$D_{J}^{\ast}(2760)$, and charmed-strange mesons $D_{s1}^{\ast}(2710)$,
$D_{sJ}^{\ast}(2860)$ and $D_{sJ}(3040)$. 
We get a quite reasonable global description of the total decay widths. A study of
the theoretical uncertainties has been performed. It consists on a montecarlo study
of the variation of the total decay widths taking into account the uncertainties of
the $\gamma$ parameters in Eq.~\ref{eq:fitgamma}.

The detailed analysis of the decay modes of every resonance is beyond the scope of
this report. However,
let us comment in more detail each sector discussing briefly the most significant
aspects.

The results predicted by the $^{3}P_{0}$ model for the well established charmed
mesons are in good agreement with the experimental data except for one case, the
total decay width of the $D^{\ast}$ meson. The $D^{\ast}$ decays only into $D\pi$
channel via strong interaction and it is assumed that the total decay width is given
mainly by this decay mode. However, the disagreement may be due to the very small
available phase space which enhances possible effects of the final-state
interactions.

With respect to the new states reported by Babar ~\refcite{PhysRevD.82.111101}, the $J^{P}=0^{-}$ is the most plausible assignment
for the $D(2550)$ meson. The total width predicted by the $^{3}P_{0}$ model with this
assignment is in very good agreement with the experimental data. The helicity-angle
distribution of $D^{\ast}(2600)$ is found to be consistent with $J^{P}=1^{-}$.
Moreover, its mass makes it the perfect candidate to be the spin partner of the
$D(2550)$ meson. The predicted mass is about $100\,{\rm MeV}$
above the experimental value while our prediction of the total decay width as the $2^{3}S_{1}$ state
agrees with the data Babar data but is in clear disagreement with the LHCb data. 

There is a strong discussion in the literature about the possible quantum
numbers that could have the mesons $D_{J}(2750)$ and $D_{J}^{\ast}(2760)$
providing a wide range of assignments. It is important to take into account the
experimental observations about these two mesons reported in
Ref.~\refcite{PhysRevD.82.111101} before assigning any quantum number. First,
despite that the two mesons are close in mass and their total widths are similar,
they are considered different particles. Second, the helicity-angle distribution
of both mesons is compatible with an angular momentum between quark and
antiquark equal to $L=2$. Third, the $D_{J}(2750)$ and $D_{J}^{\ast}(2760)$
mesons have only been seen in
the decay mode $D^{\ast}\pi$ and $D\pi$, respectively. And finally, the
following branching ratio has been measured
\begin{equation}
\frac{{\cal B}(D_{J}^{\ast}(2760)^{0} \to D^{+}\pi^{-})}
{{\cal B}(D_{J}(2750)^{0} \to D^{\ast+}\pi^{-})} = 0.42\pm0.05\pm0.11.
\label{eq:ratio2760}
\end{equation}

The assignment of the $D_{J}(2750)$ meson as the
$nJ^{P}=1\,2^{-}$ state and the $D_{J}^{\ast}(2760)$ meson as the
$nJ^{P}=1\,3^{-}$ state seems the most plausible in our model.  
This assignment agrees with those of Ref.~\refcite{PhysRevD.86.054024} 

The predicted masses are of the order of 
$100\,{\rm MeV}$ above the experimental data and we obtain a value of $0.68$ for the branching ratio written in
Eq.~(\ref{eq:ratio2760}). However the predicted widths are in clear disagreement with the experimental ones and 
the problems with the identification of these two states still remains open.

Our theoretical results are in good agreement with the experimental data in the
charmed-strange sector. Two new charmed-strange resonances, $D_{s1}(2710)$ and
$D_{sJ}(2860)$, have been observed by the BaBar Collaboration in both $DK$ and
$D^{\ast}K$ channels~\cite{PhysRevD.80.092003}. In the $D^{\ast}K$ channel, the BaBar
Collaboration has also found evidence for the $D_{sJ}(3040)$, but there is no signal
of $D_{sJ}(3040)$ in the $DK$ channel. It is commonly believed that the
$D_{s1}(2710)$ is the first excitation of the $D_{s}^{\ast}$ meson. With this
assignment, the prediction of the $^{3}P_{0}$ model is in agreement with the
experimental data. In Table~\ref{tab:totalwidths} we show the total strong decay
width of the $D_{sJ}^{\ast}(2860)$ as the third excitation of the $1^{-}$ meson and
as the ground state of the $3^{-}$ meson. The comparison between experimental data
and our results favors the $n\,J^{P}=1\,3^{-}$ assignment. The $2P$ multiplet
mean mass is predicted in our model to be near the mass of the $D_{sJ}(3040)$ resonance.
The only decay mode in which the $D_{sJ}(3040)$ has been seen until now is the
$D^{\ast}K$, and so the most possible assignment is that the $D_{sJ}(3040)$ meson
being the next excitation in the $1^{+}$ channel. Table~\ref{tab:totalwidths} shows
our prediction of the $D_{sJ}(3040)$ decay width as the $nJ^{P}=3\,1^{+}$ or
$4\,1^{+}$ state. Both are large but compatible with the experimental data.

One can see that the general trend of the total decay widths is well reproduced in
the $1^{--}$ $c\bar{c}$ sector. There are two particular cases in which the
theoretical results exceed the experimental ones. The first case is the $\psi(4415)$
resonance, where we predict a total width of $159\,{\rm MeV}$ while the PDG average
value is $62\pm20\,{\rm MeV}$~\cite{PDG2012}. However, one should mention that the
experimental data are clustered around two values ($\sim\!100\,{\rm MeV}$ and
$\sim\!50\,{\rm MeV}$) corresponding the lower one to very old measurements. If we
compare our result with the recent experimental data reported by Seth {\it et
al.}~\cite{PhysRevD.72.017501} ($\Gamma=119\pm16\,{\rm MeV}$), they are more
compatible. The second result which disagrees with the experimental data is the
corresponding to the pair of states in the vicinity of $4.6\,{\rm GeV}$. Both
widths are larger than the experimental results. The smallest total width of the
$X(4660)$ favors the $4^{3}D_{1}$ option for this state although interference
between the two states can be the origin of the disagreement.

\subsection{Description of the $D_{s1}(2536)^{+}$ strong decay properties}

Recently, new observables of the $D_{s1}(2536)^{+}$ have been measured.
The BaBar Collaboration has performed a high precision measurement of the
$D_{s1}(2536)$ decay width obtaining a value of $(1.03\pm0.05\pm0.12)\,{\rm
MeV}$~\cite{aubert2006precision}. Furthermore, the Belle Collaboration has reported
the first observation of the $D_{s1}(2536)^{+} \rightarrow D^{+}\pi^{-}K^{+}$ decay
measuring the branching fraction~\cite{PhysRevD.77.032001}
\begin{equation}
\frac{D_{s1}(2536)^{+} \to D^{+}\pi^{-}K^{+}}{D_{s1}(2536)^{+} \to
D^{\ast+}K^{0}}=(3.27\pm0.18\pm0.37)\%.
\end{equation}
They also measured the ratio of the $S$-wave amplitude in the $D_{s1}(2536)^{+}
\rightarrow D^{\ast+}K^0$ decay finding a value of $0.72\pm0.05\pm0.01$.

In order to gain insight into the structure of the $1^{+}$ charmed-strange
mesons,
we study the reaction $D_{s1}(2536)^{+} \rightarrow D^{+}\pi^{-}K^{+}$ as well
as the angular decomposition of the $D_{s1}(2536)^{+}\rightarrow D^{\ast+}K^{0}$
decay.

In the model described in this work, a tetraquark $c\bar{s}n\bar{n}$ state has been
predicted by Vijande {\it et al.} in Ref.~\refcite{PhysRevD.73.034002} with quantum
numbers $IJ^{P}=0\,1^{+}$ and mass $M=2841\,{\rm MeV}$. If this state is present it
should be coupled to the $J^{P}=1^{+}$ $c\bar{s}$ states.

Working in the HQS limit, the $c\bar{s}n\bar{n}$ tetraquark has three different spin
states, $|0\,1/2\rangle$, $|1\,1/2\rangle$ and $|1\,3/2\rangle$ where the first index
denotes the spin of the $n\bar{n}$ pair and the second the coupling with the
$\bar{s}$ spin. Although we use the $^{3}P_{0}$ model to calculate the meson decay
widths, a description of the coupling between the $D_{s}$ meson and the tetraquark
based on this model is beyond the scope of the present calculation. However, we use
it here to select the dominant couplings and parametrize the vertex as a constant
$C_S$. The model assumes that the $n\bar{n}$ pair created is in a $J=0$ state which
means that the $D_s$ states will only couple with the first tetraquark component
which has spin $1/2$ for the three light quarks. In the HQS limit the heavy quark is
an spectator and the angular momentum of the light quarks has to be conserved so that
the tetraquark will only couple to the $c\bar{s}$ $j_q=1/2$ state.

For that reason we couple the tetraquark structure with the $j_q=1/2$ $c\bar{s}$
state. This choice differs from the one performed in
Ref.~\refcite{PhysRevD.73.034002} where the tetraquark is only coupled to the
$^{1}P_{1}$ state and not to the $^{3}P_{1}$. However, this choice has several
advantages: it has the correct heavy quark limit, it may reproduce the narrow
width of the $D_{s1}(2536)^{+}$ state and it is in agreement with the experimental
situation which tells us that the prediction of the heavy quark limit is reasonable
for the $j_q=3/2$ state but not for the $j_q=1/2$ one.

In this scenario we diagonalize the matrix
\begin{equation}
M = \left( \begin{array}{ccc} 
M_{^3P_1}            & C_{SO}               & \sqrt{\frac{2}{3}}\,C_S \\
C_{SO}               & M_{^1P_1}            & \sqrt{\frac{1}{3}}\,C_S \\
\sqrt{\frac{2}{3}}\,C_{S} & \sqrt{\frac{1}{3}}\,C_{S} & M_{c\bar{s}n\bar{n}}
\end{array} \right),
\end{equation}
where $M_{^{3}P_{1}}=2571.5\,{\rm MeV}$, $M_{^{1}P_{1}}=2576.0\,{\rm MeV}$ and 
$M_{c\bar{s}n\bar{n}}=2841\,{\rm MeV}$ are the masses of the states without
couplings, the $C_{SO}=19.6\,{\rm MeV}$ is the coupling induced by the antisymmetric
spin-orbit interaction calculated within the model and $C_S$ is the parameter that
gives the coupling between the $j_q=1/2$ component of the $^{3}P_{1}$ and $^{1}P_{1}$
states and the tetraquark. The value of the parameter $C_S=224\,{\rm MeV}$ is fitted
to the mass of the $D_{s1}(2460)$. We get the three eigenstates shown in
Table~\ref{tab:wfDs1p}. There we also show the probabilities of the three components
for each state and the relative phases between different components. One can see
that the $D_{s1}(2460)$ meson has a sizable non-$q\bar{q}$ component whereas the
$D_{s1}(2536)$ is almost a pure $q\bar{q}$ state. The presence of non-$q\bar{q}$
degrees of freedom in the $1^{+}$ $c\bar{s}$ channel enhances the $j_{q}=3/2$
component of the $D_{s1}(2536)$ meson. Moreover, a $1^+$ state with an important
component of $c\bar{s}n\bar{n}$ tetraquark structure is found at $2973\,{\rm MeV}$.

\begin{table}[!t]
\begin{center}
\caption{\label{tab:wfDs1p} Masses and probability distributions for the three
eigenstates obtained from the coupling of the $D_{s}$ and tetraquark states. The
relative sign to the tetraquark component is also shown.}
\begin{tabular}{ccccccc}
\toprule
$M$ (MeV) & $S(^{3}P_{1})$ & $P(^{3}P_{1})$ & $S(^{1}P_{1})$ & $P(^{1}P_{1})$
& $S(c\bar{s}n\bar{n})$ & $P(c\bar{s}n\bar{n})$ \\
\colrule
$2459$ & $-$ & $55.7$ & $-$ & $18.8$ & $+$ & $25.5$ \\
$2557$ & $+$ & $27.7$ & $-$ & $72.1$ & $+$ & $0.2$ \\
$2973$ & $+$ & $16.6$ & $+$ & $9.1$ & $+$ & $74.3$ \\
\botrule
\end{tabular}
\end{center}
\end{table}

\begin{table}[!t]
\begin{center}
\caption {\label{tab:propertiesDs1} Width and the three branching ratios defined in
the text. The first row shows the experimental data and the second shows our results
for the $D_{s1}(2536)$ state given in Table~\ref{tab:wfDs1p}. For completeness
we give in the last two rows the results for the two $1^{+}$ $c\bar{s}$ states
predicted by the naive CQM.}
\scalebox{0.90}{\begin{tabular}{ccccc}
\toprule
$M$ (MeV) & $\Gamma$ (MeV) & $R_1$ & $R_2$ & $R_3(\%)$ \\
\colrule
Exp. & $1.03\pm0.05\pm0.12$ & $1.27\pm0.21$ & $0.72\pm0.05\pm0.01$ &
$3.27\pm0.18\pm0.37$ \\
\colrule
$2557$ & $0.99$ & $1.31$ & $0.66$ & $14.07$ \\[2ex]
$2593$ & $190.17$ & $1.09$ & $1.00$ & $13.13$ \\
$2554$ & $11.24$ & $1.11$ & $0.97$ & $13.19$ \\
\botrule
\end{tabular}}
\end{center}
\end{table}

We now calculate the different decay widths for the $D_{s1}(2536)^{+}$ state of
Table~\ref{tab:wfDs1p}. As expected, the $D^{\ast}K$ decay width is narrow
$\Gamma=0.99\,{\rm MeV}$. As the $DK$ decay is suppressed, the total width would be
mainly given by the $D^{\ast}K$ channel and is in the order of the experimental value
$\Gamma_{\rm exp}=(1.03\pm0.05\pm0.12)\,{\rm MeV}$ measured by
BaBar~\cite{aubert2006precision}. Of course the value strongly depends on the
$^{3}P_{0}$ $\gamma$ strength parameter that has been determined before by a global
fit of the total decay widths of heavy mesons. It also depends on the fact that we
have only coupled the $1/2$ state with the tetraquark making the remaining state a
purest $3/2$ which makes it narrower. If we would include an small coupling between
the $3/2$ state and the tetraquark our $D_{s1}(2536)$ will be broader.

There are two other experimental data that do not depend on the $\gamma$ parameter,
namely the branching ratio~\cite{PDG2012}
\begin{equation}
R_{1} = \frac{\Gamma(D_{s1}(2536)^{+} \rightarrow D^{\ast0}K^{+})}
{\Gamma(D_{s1}(2536)^{+} \rightarrow D^{\ast+}K^{0})} = 1.27\pm0.21,
\end{equation}
and the ratio of $S$-wave over the full width for the $D^{\ast+}K^{0}$
decay~\cite{PhysRevD.77.032001}
\begin{equation}
R_{2} = \frac{\Gamma_{S}(D_{s1}(2536)^{+} \rightarrow D^{\ast+}K^{0})}
{\Gamma(D_{s1}(2536)^{+} \rightarrow D^{\ast+}K^{0})} = 0.72\pm0.05\pm0.01.
\end{equation}
The first branching ratio should be $1$ if the isospin symmetry was exact. However,
the charge symmetry breaking in the phase space makes it different from this value.
The effect is sizable since the $D_{s1}(2536)^{+}$ is close to the $D^{\ast}K$
threshold and for this reason it also depends on the details of the $D_{s1}$ wave
function. We get for this ratio the value $R_{1}=1.31$, in good agreement with the
experimental one.

Notice that in order to get $R_{2}$ different from one, we need to have a state
with high $j_{q}=3/2$ component. In our case we get a value of $R_{2}=0.66$,
close to the experimental data. The fact that our result is smaller than the
experimental one indicates that the probability of the $j_{q}=3/2$ state is high
which is in agreement with the fact that we get a narrower state.

Finally we calculate the branching
\begin{equation}
R_{3} = \frac{\Gamma(D_{s1}(2536)^{+}\rightarrow D^{+}\pi^{-}K^{+})}
{\Gamma(D_{s1}(2536)^{+}\rightarrow D^{*+}K^{0})} = (3.27\pm 0.18\pm0.37)\%.
\end{equation}
As the $D^{+}\pi^{-}$ pair in the final state is the only $D\pi$ combination that
cannot come from a $D^{\ast}$ resonance, we describe the reaction through a virtual
$D^{\ast0}$ meson since $M_{D^{\ast0}} < M_{D^{+}}+M_{\pi^{-}}$. We get $R_3=14.1\%$,
a factor $3-4$ greater than the experimental data. This value seems not to depend on
the details of the $D_{s1}$ wave function.

All these results for the width and the ratios $R_{1}$, $R_{2}$ and $R_{3}$ are
summarized in Table~\ref{tab:propertiesDs1}. We also show, for the sake of
completeness, the results for the two $1^{+}$ states without coupling to the
$c\bar{s}n\bar{n}$ tetraquark. None of these two states agree with the full
set of experimental values.

%%%%%%%%%%%%%%%%%%%%%%%%%%%%%%%%%%%%%%%%%%%%%%%%%%%%%%%%%%%%%%%%%%%%%%%%%%%%%%%%
\FloatBarrier
%%%%%%%%%%%%%%%%%%%%%%%%%%%%%%%%%%%%%%%%%%%%%%%%%%%%%%%%%%%%%%%%%%%%%%%%%%%%%%%%

\section{Charmonium resonances in $e^{+}e^{-}$ exclusive reactions around the
$\psi(4415)$ region}

The Belle Collaboration has recently performed measurements of the exclusive
cross section for the processes $e^{+}e^{-}\rightarrow
D^{0}D^{-}\pi^{+}$~\cite{PhysRevLett.100.062001} and $e^{+}e^{-}\rightarrow
D^{0}D^{\ast-}\pi^{+}$~\cite{PhysRevD.80.091101} over the center-of-mass energy range
$4.0\,{\rm GeV}$ to $5.0\,{\rm GeV}$. In the first reaction they found a prominent
peak in the cross section which is interpreted as the $\psi(4415)$. From the study of
the resonant structure in the $\psi(4415)$ decay, they conclude that the final
channel $D^{0}D^{-}\pi^{+}$ is reached through the
$D\bar{D}_{2}^{\ast}(2460)$ intermediate state. Using a relativistic
Breit-Wigner function parametrization, they obtain the value of the
$\mathcal{B}(\psi(4415)\rightarrow D\bar{D}_{2}^{\ast}(2460))\times\mathcal{B}
(\bar{D}_{2}^{\ast}(2460)\rightarrow D\pi^{+})$ product of branching fractions
and the mass and width of the $\psi(4415)$. From the measurement of the
$e^{+}e^{-}\rightarrow D^{0}D^{\ast-}\pi^{+}$ exclusive cross section reported
in Ref.~\refcite{PhysRevD.80.091101}, they provide upper limits on the peak cross
section for the process $e^{+}e^{-}\rightarrow X \rightarrow
D^{0}D^{\ast-}\pi^{+}$ where $X$ denotes $X(4260)$, $X(4360)$, $\psi(4415)$,
$X(4630)$ and $X(4660)$. Although only the value concerning the $\psi(4415)$ is
significant.

We have seen that our assignment of the $\psi(4415)$ as a $D$-wave state leaving
the $4S$ state for the $X(4360)$ agrees with the last measurements of the
leptonic and total decay widths. Now we want to perform a study of the two above
reactions to test if our result is also compatible with the measurements of Belle.

We assume the reaction $e^{+}e^{-}\rightarrow X\rightarrow DD^{(\ast)}\pi$ and
parametrize the cross section using a relativistic Breit-Wigner function
including Blatt-Weisskopf corrections. The relativistic Breit-Wigner amplitude
for the process ``$e^{+}e^{-}\rightarrow \mbox{\it resonance} \rightarrow
\mbox{\it hadronic final state f}\,$'' at center-of-mass energy $\sqrt{S}$ can be
written as
\begin{equation}
\mathcal{T}_{r}^{f}(\sqrt{S})=\frac{M_{r}\sqrt{\Gamma_{r}^{ee}\Gamma_{r}^{f}}}{
S-M^{2}_{r}+iM_{r}\Gamma_{r}}e^{i\delta_{r}},
\end{equation}
where $r$ indicates the resonance being studied, $M_{r}$ is the nominal mass,
$\Gamma_{r}$ is the full width, $\Gamma^{ee}_{r}$ is the leptonic width,
$\Gamma^{f}_{r}$ is the hadronic width for the decaying channel $f$ and
$\delta_{r}$ is a relative phase.

When there are more than one resonance in the same energy range and we measure
the same decay channel, the spin-averaged cross section is a coherent sum of the
Breit-Wigner amplitudes for each resonance
\begin{equation}
\sigma(\sqrt{S})=\frac{(2J+1)}{(2S_{1}+1)(2S_{2}+1)}\frac{16\pi}{S}
\left|\sum_{r}\frac{M_r\sqrt{\Gamma_{r}^{ee}\Gamma_{r}^{f}}}{S-M_{r}^{2}
+iM_{r} \Gamma_{r}}e^{i\delta_{r}}\right|^{2}.
\label{eq:crosssection}
\end{equation}

Now, we introduce the energy dependence of the widths following
Ref.~\refcite{Ablikim2008315}. The angular momentum dominant partial width of a
resonance decaying into one channel is given by~\cite{Barbaro-Galtieri}
\begin{equation}
\Gamma_{r}^{f}(\sqrt{S})=\hat{\Gamma}_{r}\frac{Z_{f}^{2L+1}}{B_{L}},
\end{equation}
with $Z_{f}$ defined as $Z_{f}\equiv\rho P_{f}$, where $P_{f}$ is the decay
momentum and $\rho$ is a free parameter whose value is around the range of the
interaction, in the order of a few fermis. The energy-dependent partial wave
functions $B_{L}(Z_{f})$ are given in either Ref.~\refcite{Barbaro-Galtieri}
or~\refcite{blatt1991theoretical}
\begin{equation}
\begin{split}
& B_{0}=1, \\
& B_{1}=1+Z_{f}^{2}, \\
& B_{2}=9+3Z_{f}^{2}+Z_{f}^{4}, \\
& B_{3}=225+45Z_{f}^{2}+6Z_{f}^{4}+Z_{f}^{6},
\end{split}
\end{equation}
and $\hat{\Gamma}_{r}$ is related with the partial width at the mass of the
resonance, $\Gamma_{0}$, as
\begin{equation}
\hat{\Gamma}_{r}=\Gamma_{0}\frac{B_{L}(P_{0})}{Z_{f}^{2L+1}(P_{0})}.
\end{equation}
Then, our final expressions for the partial and total width are given by
\begin{equation}
\begin{split}
&
\Gamma_{r}^{f}(\sqrt{S})=\Gamma_{0}\frac{Z_{f}^{2L+1}(P_{f})}{Z_{f}^{2L+1}(P_{0}
)}\frac{B_{L}(P_{0})}{B_{L}(P_{f})}, \\
&
\Gamma_{r}(\sqrt{S})=\frac{2M_{r}}{M_{r}+\sqrt{S}}\sum_{f}\Gamma_{r}^{f}
(\sqrt{S}),
\end{split}
\end{equation}
where the term $\frac{2M_{r}}{M_{r}+\sqrt{S}}$ is a relativistic correction
factor~\cite{Barbaro-Galtieri}.

\subsection{The process $e^{+}e^{-}\rightarrow D^{0}D^{-}\pi^{+}$}

This process has been studied by Pakhlova {\it et al.} in
Ref.~\refcite{PhysRevLett.100.062001}. They perform a separate study of the
$e^{+}e^{-}\rightarrow D\bar{D}_{2}^{\ast}(2460)$ and $e^{+}e^{-}\rightarrow
D(D\pi)_{{\rm non-}\bar{D}_{2}^{\ast}(2460)}$ concluding that the
$e^{+}e^{-}\rightarrow D^{0}D^{-}\pi^{+}$ is dominated by $X\rightarrow
D\bar{D}_{2}^{\ast}(2460)$.

Assuming $X\equiv\psi(4415)$ and a relativistic Breit-Wigner function to fit
the data, the peak cross section for the process $e^{+}e^{-}\rightarrow
X\rightarrow D\bar{D}_{2}^{\ast}(2460)$ is
$\sigma(e^{+}e^{-}\rightarrow\psi(4415))\times \mathcal{B}(\psi(4415)\rightarrow
D\bar{D}_{2}^{\ast}(2460))\times \mathcal{B}(\bar{D}_{2}^{\ast}(2460)\rightarrow
D\pi^{+})=(0.74\pm0.17\pm0.08)\,{\rm nb}$.

Using
\begin{equation}
\sigma(e^{+}e^{-}\rightarrow
X)=\frac{12\pi}{m_{X}^{2}}\frac{\Gamma_{ee}}{\Gamma_{\rm tot}},
\end{equation}
the authors of Ref.~\refcite{PhysRevLett.100.062001} estimate
$\mathcal{B}(\psi(4415)\rightarrow D\bar{D}_{2}^{\ast}(2460))\times\mathcal{B}
(\bar{D}_{2}^{\ast}(2460)\rightarrow D\pi^{+})=(10.5\pm2.4\pm3.8)\%$ or
$(19.5\pm4.5\pm9.2)\%$ depending on the different parametrization of the
$\psi(4415)$ resonance (Refs.~\refcite{PDG2012} and~\refcite{Ablikim2008315},
respectively).

Furthermore, taken from Ref.~\refcite{PDG2012} the branching fraction for
$\bar{D}_{2}^{\ast}(2460)\rightarrow D\pi^{+}$, one can estimate
$\mathcal{B}(\psi(4415)\rightarrow DD_{2}^{\ast})=0.47$ using the
resonance parameters of Ref.~\refcite{PDG2012} or $0.86$
using those of Ref.~\refcite{Ablikim2008315}. Note that there are two final charged
states in the calculation of $\mathcal{B} (\bar{D}_{2}^{\ast}(2460)\rightarrow
D\pi^{+})$ and we give the branching fraction of the process $\psi(4415)\to
DD_{2}^{\ast}$ in function of the $DD_{2}^{\ast}$ state and not in function of
the $D\bar{D}_{2}^{\ast}$ one.

The theoretical calculation of the $e^{+}e^{-}\rightarrow D^{0}D^{-}\pi^{+}$
cross section can be divided in three steps. The first one is the resonance
production $e^{+}e^{-}\rightarrow X$ which can be given in terms of the leptonic
width. The second and third steps are the strong decays $\psi(4415)\rightarrow
D\bar{D}_{2}^{\ast}(2460)$ and $\bar{D}_{2}^{\ast}(2460)\rightarrow D\pi^{+}$
which can be calculated using the $^{3}P_{0}$ model. These two partial widths
are involved in the calculation of the $\Gamma_{r}^{f}$ in
Eq.~(\ref{eq:crosssection}) because in the case under study we have
$\Gamma_{r}^{f}=\Gamma(X\equiv\psi(4415)\rightarrow D\bar{D}_{2}^{\ast}
(2460)\rightarrow DD\pi^{+})$ which is equal to $\Gamma(X\equiv
\psi(4415)\rightarrow D\bar{D}_{2}^{\ast}(2460))\times\mathcal{B}(\bar{
D}_{2}^{\ast}(2460)\rightarrow D\pi^{+})$.

\begin{table}[!t]
\begin{center}
\caption {\label{tab:resonances} Resonance parameters predicted by our constituent
quark model for the $X(4360)$ and $\psi(4415)$. The experimental data are taken from
Ref.~\protect\refcite{PDG2012} for $X(4360)$ and
Ref.~\protect\refcite{Ablikim2008315} for
$\psi(4415)$.}
\begin{tabular}{lcccc}
\toprule
& \multicolumn{2}{c}{$X(4360)$} & \multicolumn{2}{c}{$\psi(4415)$} \\
& Theory & Experiment & Theory & Experiment \\
\colrule
Mass (MeV) & $4389$ & $4361\pm9\pm9$ & $4426$ & $4415.1\pm7.9$ \\
$\Gamma_{\rm tot}\,\mbox{(MeV)}$ & $113.9$ & $74\pm15\pm10$ & $159.0$ &
$71.5\pm19.0$ \\
$\Gamma_{ee}\,\mbox{(keV)}$ & $0.78$ & - & $0.33$ & $0.35\pm0.12$ \\
\botrule
\end{tabular}
\end{center}
\end{table}

We show the prediction of our model for the mass, the total width and the
leptonic width of the resonance $\psi(4415)$ in Table~\ref{tab:resonances}.
First, we calculate the branching fractions
\begin{equation} 
\begin{split}
& \mathcal{B}(D_{2}^{\ast+}\rightarrow D^{0}\pi^{+})=0.43
~(\mbox{Exp.:}~0.44\pm0.09), \\
& \mathcal{B}(\bar{D}_{2}^{\ast0}\rightarrow D^{-}\pi^{+})=0.43
~(\mbox{Exp.:}~0.47\pm0.03),
\end{split}
\end{equation}
which agree with the experimental values of Ref.~\refcite{PDG2012}. Furthermore the
ratios
\begin{equation} 
\begin{split}
& 
R_1=\frac{\Gamma(D_{2}^{\ast+}\rightarrow D^{0}\pi^{+})}{\Gamma(D_{2}^{\ast+}
\rightarrow D^{*0}\pi^{+})}=1.81~(\mbox{Exp.:}~1.9\pm 1.1\pm 0.3), \\
& 
R_2=\frac{\Gamma(\bar{D}_{2}^{\ast0}\rightarrow
D^{-}\pi^{+})}{\Gamma(\bar{D}_{2} ^{\ast0}\rightarrow D^{\ast-}\pi^{+})}
=1.81~(\mbox{Exp.:}~1.56\pm 0.16),
\end{split}
\end{equation}
also agree with the experimental data of Ref.~\refcite{PDG2012}.
 
However, when in a similar way we calculate the $\mathcal{B}(\psi(4415)\rightarrow
DD_{2}^{\ast})$, we obtain $0.15$ which clearly disagrees with the estimation of
Ref.~\refcite{PhysRevLett.100.062001}.

\begin{figure}[!t]
\begin{center}
\parbox[c]{0.49\textwidth}{
\centering
\includegraphics[width=0.42\textwidth]{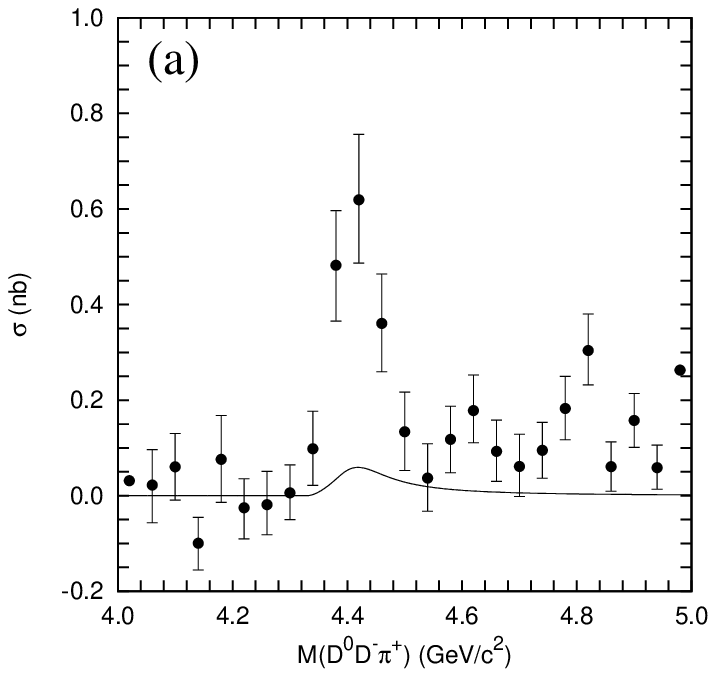}
}
\parbox[c]{0.49\textwidth}{
\centering
\includegraphics[width=0.42\textwidth]{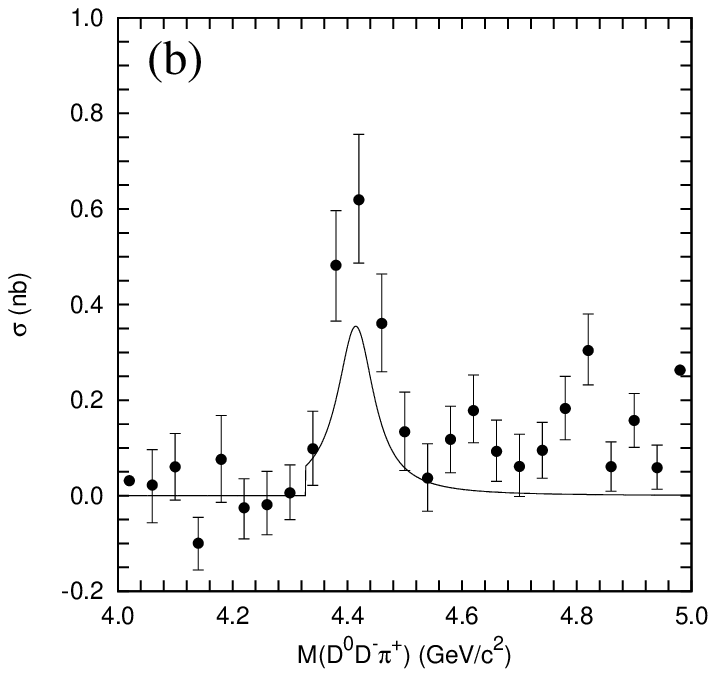}
}
\parbox[c]{0.49\textwidth}{
\centering
\includegraphics[width=0.42\textwidth]{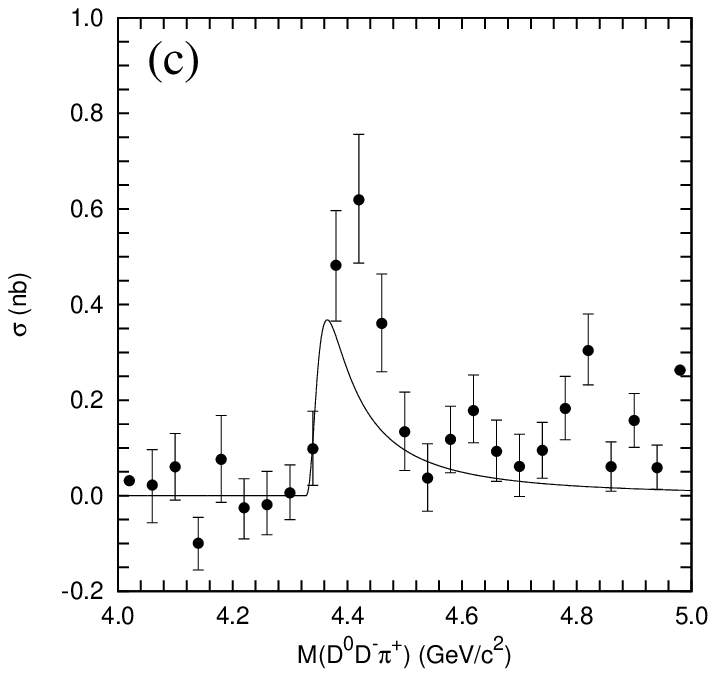}
}
\parbox[c]{0.49\textwidth}{
\centering
\includegraphics[width=0.42\textwidth]{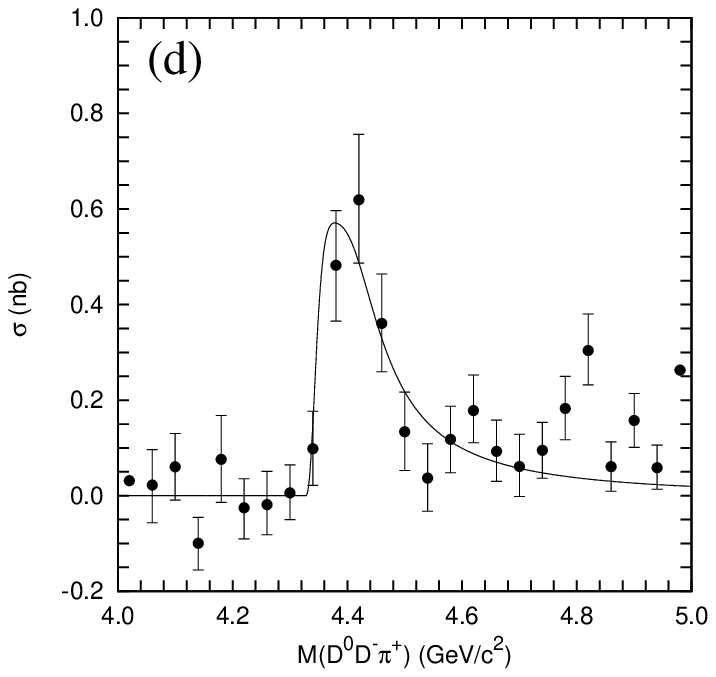}
}
\caption{\label{fig:DDpi} {\bf (a)}: Our model prediction with only
the resonance $\psi(4415)$. {\bf (b)}: Model prediction of
Ref.~\protect\refcite{PhysRevD.72.054026}. {\bf (c)}: Our model prediction with only
the resonance $X(4360)$. {\bf (d)}: Our model prediction with the interference of
the resonances $X(4360)$ and $\psi(4415)$.}
\end{center}
\end{figure}

Our model prediction for the cross section is shown in panel (a) of
Fig.~\ref{fig:DDpi}. One can see that our result is very far from the experimental
data. In order to test if this disagreement is due to the $3D$ character of our
resonance, we repeat the calculation using the parametrization of
Ref.~\refcite{PhysRevD.72.054026} where the $\psi(4415)$ is described as a $4S$
state. Although the result approaches the experimental data, see
Fig.~\ref{fig:DDpi}(b), it still does not describe the full cross section. Certainly,
the theoretical results have some uncertainties coming either from the wave functions
used in the $^{3}P_{0}$ model or the leptonic width. To minimized these uncertainties
we have used in Fig.~\ref{fig:DDpi}(b) the experimental value for the leptonic
width~\cite{PDG2012}. Using the value $\Gamma_{e^{+}e^{-}}$ predicted by the model of
Ref.~\refcite{PhysRevD.72.054026}, the result would be a factor $\sim\!3$ smaller.

Taken into account that the energy window around the nominal $\psi(4415)$ mass
in the experiment of Ref.~\refcite{PhysRevLett.100.062001} is $\pm100\,{\rm MeV}$,
we introduce in the calculation the resonance $X(4360)$ which appears as a $4S$
$1^{--}$ $c\bar{c}$ meson in our model. The predicted mass, total and leptonic
widths are shown in Table~\ref{tab:resonances}. Panel (c) of Fig.~\ref{fig:DDpi}
shows how this resonance alone cannot reproduce the data but the interference
between the $X(4360)$ and $\psi(4415)$, panel (d) of Fig.~\ref{fig:DDpi},
produces a remarkable agreement with the data.

Using the interference of the two resonances, the theoretical value for the
exclusive cross section $\sigma(e^{+}e^{-}\rightarrow
D\bar{D}_{2}^{\ast}(2460)\rightarrow D^{0}D^{-}\pi^{+})$ at the $\psi(4415)$
mass is $0.48\,{\rm nb}$, within the error bars of the experimental one:
$(0.62^{+0.14}_{-0.13})\,{\rm nb}$. Our result indicates that the two resonances
are needed to explain the experimental data.

\subsection{The process $e^{+}e^{-}\rightarrow D^{0}D^{\ast-}\pi^{+}$}

Using the same philosophy we check the $e^{+}e^{-}\rightarrow D^{0}D^{\ast-}\pi^{+}$
exclusive cross section measured by the Belle
Collaboration~\cite{PhysRevD.80.091101}. The experimental analysis estimates from the
amplitude of a relativistic Breit-Wigner function fitted to the data an upper limit
of $0.76\,{\rm nb}$ for the peak cross section at $E_{\rm cm}=M_{\psi(4415)}$. 

We calculate the cross section following the same procedure as before. Again the
resonance production $e^{+}e^{-}\rightarrow X$ has been calculated and is given in
Table~\ref{tab:resonances}. Now, the second and third steps are the strong decays
$\psi(4415)\rightarrow D^{\ast-}D^{\ast+}$ and $D^{\ast+}\rightarrow D^{0}\pi^{+}$.

The theoretical result for the branching fraction $\mathcal{B}(D^{\ast+} \rightarrow
D^{0}\pi^{+})$ is $0.687$, in very good agreement with the experimental value
$0.677\pm0.006$ of Ref.~\refcite{PDG2012}. For the other branching fraction,
$\mathcal{B}(\psi(4415)\rightarrow D^{\ast}D^{\ast})$, there is no experimental data.
Our theoretical result is $0.20$.

\begin{figure}[!t]
\begin{center}
\parbox[c]{0.49\textwidth}{
\centering
\includegraphics[width=0.42\textwidth]{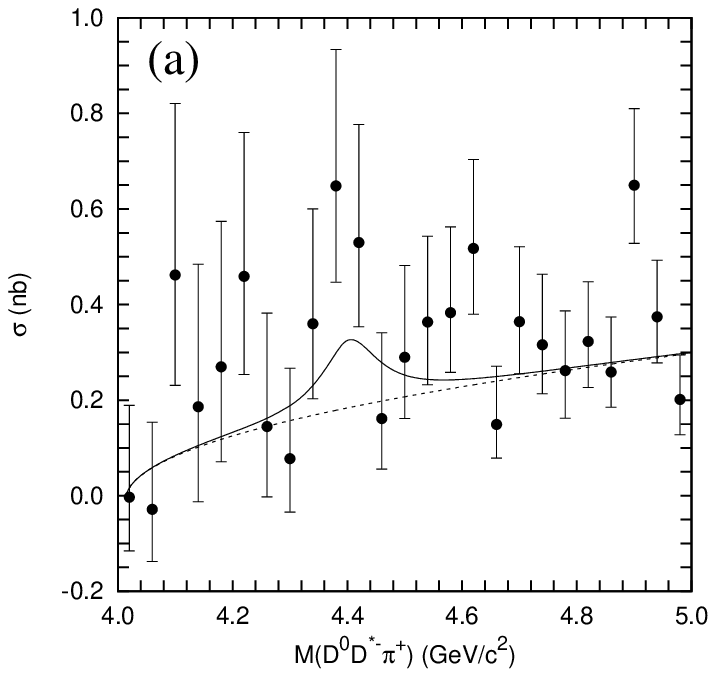}
}
\parbox[c]{0.49\textwidth}{
\centering
\includegraphics[width=0.42\textwidth]{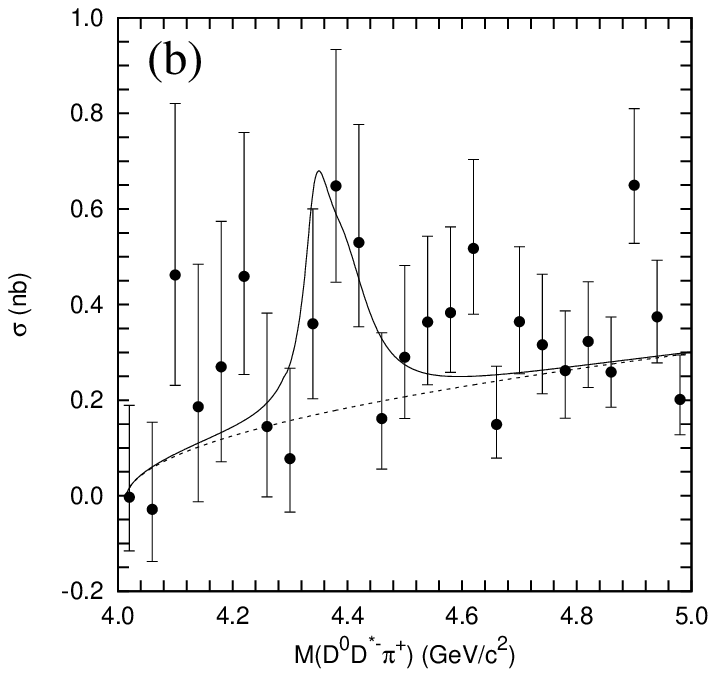}
}
\caption{{\bf (a)}: Our model prediction with only the resonance
$\psi(4415)$. {\bf (b)}: Our model prediction with the interference of the
resonances $X(4360)$ and $\psi(4415)$.}
\label{fig:DDastpi}
\end{center}
\end{figure}

The calculation of the cross section including the $\psi(3D)$ resonance with
$M=4426\,{\rm MeV}$ alone does not reproduce the full strength of the resonance
at $E_{\rm cm}=M_{\psi(4415)}$ and the result is improved when the $X(4360)$ is
added. See Fig.~\ref{fig:DDastpi}.

From the cross section of Fig.~\ref{fig:DDastpi}(b), we calculate the peak cross
section for the $e^{+}e^{-}\rightarrow D^{0}D^{\ast-}\pi^{+}$ process at
$M(D^{0}D^{\ast-}\pi^{+})=4415\,{\rm MeV}$ obtaining $0.45\,{\rm nb}$, which is
compatible with the experimental upper limit $0.76\,{\rm nb}$ at $90\%$ C.L. This
result is also compatible with the upper limits measured in
Ref.~\refcite{PhysRevD.80.091101} for the branchings ${\mathcal B}_{ee} \times
{\mathcal B}(X\to D^{0}D^{\ast-}\pi^{+})$ where $X$ denotes the $X(4360)$ and
$\psi(4415)$. We obtain the value $0.25\times10^{-6}$ for the $X(4360)$ and
$0.35\times10^{-6}$ for the $\psi(4415)$. They should be compared with the upper
limits $<\!0.72\times10^{-6}$ and $<\!0.99\times10^{-6}$, respectively.

The $X(4360)$ resonance has been sometimes assigned as an unconventional charmonium
state since it was discovered in the $e^{+}e^{-}\to \pi^{+}\pi^{-}\psi(2S)$
decay~\cite{PhysRevLett.99.142002} and its open-charm decays were assumed to be
suppressed. Ref.~\refcite{PhysRevD.80.091101} gives the branching ratio
$\mathcal{B}(X\to D^{0}D^{\ast-}\pi^{+})/\mathcal{B}(X\to\pi^{+}\pi^{-}
\psi(2S))<8$. Since the $X\to \pi^+\pi^-\psi(2S)$ is an Okubo-Zweig-Iizuka (OZI)
suppressed decay the value of this upper limit means that the open-charm
$D^{\ast+}D^{\ast-}$ decay, where $D^{\ast+}$ decays into $D^{0}\pi^{+}$, should be
small. This is actually the case in our model. We get
$\Gamma(D^{0}D^{\ast-}\pi^{+})=\Gamma(X(4360)\to D^{\ast+}D^{\ast-})\mathcal{B}(D^{
\ast+}\to D^{0}\pi^{+})=3.0\,{\rm MeV}$ and combined with the experimental
information we obtain $\Gamma(X(4360)\to \psi(2S)\pi^{+}\pi^{-})\gtrsim375\,{\rm
keV}$ which is in the same order of magnitude that other similar decays. The decay of
the $\psi(2S)$ meson into $J/\psi\pi\pi$ has a width of $147\,{\rm keV}$ according to
PDG~\cite{PDG2012}.

\begin{figure}[!t]
\begin{center}
\parbox[c]{0.49\textwidth}{
\centering
\includegraphics[width=0.42\textwidth]{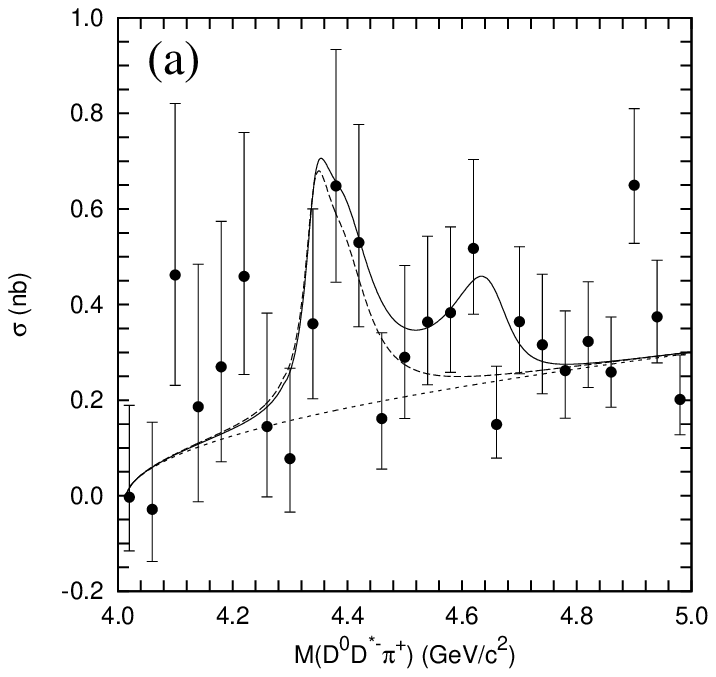}
}
\parbox[c]{0.49\textwidth}{
\centering
\includegraphics[width=0.42\textwidth]{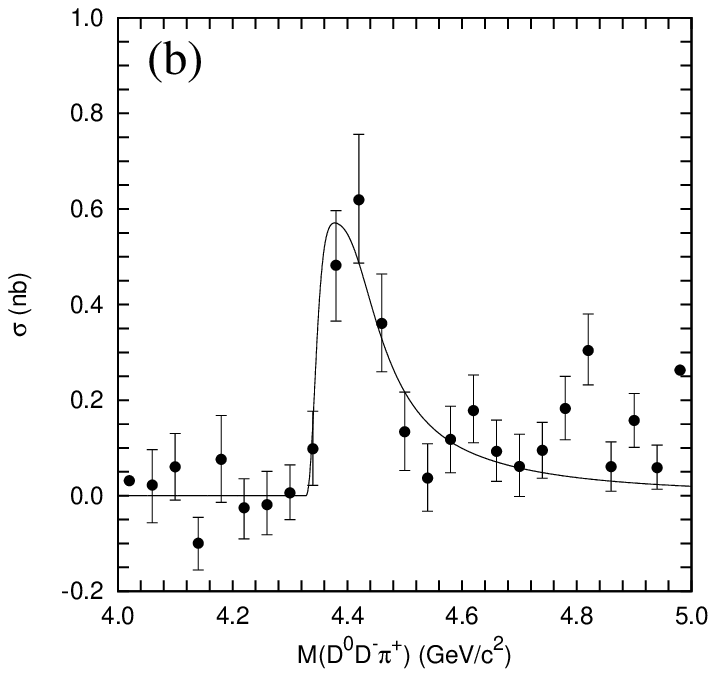}
}
\caption{\label{fig:NR1} {\bf (a)}: Model prediction of the reaction
$e^{+}e^{-}\rightarrow D^{0}D^{\ast-}\pi^{+}$ with the resonances $X(4360)$ and
$\psi(4415)$ (dashed line) and including $\psi(5S)$ and $\psi(4D)$ (solid line).
{\bf (b)}: Model prediction of the reaction $e^{+}e^{-}\rightarrow
D^{0}D^{-}\pi^{+}$ with the resonances $X(4360)$ and $\psi(4415)$ (dashed line)
and including $\psi(5S)$ and $\psi(4D)$ (solid line).}
\end{center}
\end{figure}

Finally, data of Ref.~\refcite{PhysRevD.80.091101} show a bump around $4.6\,{\rm
GeV}$ although data of Ref.~\refcite{PhysRevLett.100.062001} do not show this bump.
Our model predicts two states $\psi(5S)$ and $\psi(4D)$ in this energy region. The
inclusion of these two resonances improves the agreement with the cross section in
the bump region, as we can see in panel (a) of Fig~\ref{fig:NR1}. This bump should
not clearly appear in the $e^{+}e^{-}\to D^{0}D^{-}\pi^{+}$, as one can see in panel
(b) of Fig.~\ref{fig:NR1}, due to the negligible decay width, predicted by the
$^{3}P_{0}$ model, of the $\psi(5S)$ and $\psi(4D)$ states into $DD_{2}^{\ast}(2460)$
channel.

%%%%%%%%%%%%%%%%%%%%%%%%%%%%%%%%%%%%%%%%%%%%%%%%%%%%%%%%%%%%%%%%%%%%%%%%%%%%%%%%
\FloatBarrier
%%%%%%%%%%%%%%%%%%%%%%%%%%%%%%%%%%%%%%%%%%%%%%%%%%%%%%%%%%%%%%%%%%%%%%%%%%%%%%%%

\section{Weak Decays}
\label{sec:WeakDecays}

$B$-factories
have become an important source of data on heavy hadrons.
Bottomonium states decay mainly into $B\bar{B}$ pairs, and these $B$ mesons
decay subsequently into charmed and charmless hadrons via the weak interaction. 

To describe theoretically the properties of the mentioned $c$-quark mesons
(conventional or unexpected), one must deal with weak interaction observables which
are generally concerned with the semileptonic and nonleptonic decays of $b$-hadrons. We
perform in this Section a study of the semileptonic and nonleptonic $B$ decays into
orbitally excited charmed and charmed-strange mesons in order to gain insight on the
structure of the charmed mesons.

\subsection{Semileptonic $B$ $(B_{s})$ decays into $D^{\ast\ast}$
$(D_{s}^{\ast\ast})$ mesons}

Different Collaborations have recently reported semileptonic $B$ decays into
orbitally excited charmed mesons providing detailed results of branching fractions.
The theoretical analysis of these data, which include both weak and strong decays,
offers the possibility for a stringent test of meson models. 

The Belle Collaboration reported data~\cite{PhysRevD.77.091503} on the product of
branching fractions $\mathcal{B}(B^{+} \to D^{\ast\ast}l^{+}\nu_{l})$ 
$\mathcal{B}(D^{\ast\ast} \to D^{(\ast)}\pi)$, where, in the usual notation, $l$
stands for a light $e$ or $\mu$ lepton. The $D_{0}^{\ast}(2400)$, $D_{1}(2430)$,
$D_{1}(2420)$ and $D_{2}^{\ast}(2460)$ mesons are denoted generically as
$D^{\ast\ast}$, and the $D^{\ast}$ and $D$ mesons as $D^{(\ast)}$.

$D^{\ast\ast}$ decays are reconstructed in the decay chains $D^{\ast\ast} \to
D^{\ast}\pi^{\pm}$ and $D^{\ast\ast} \to D\pi^{\pm}$. In particular, the
$D_{0}^{\ast}(2400)$ meson decays only through the $D\pi$ channel, while the
$D_{1}(2430)$ and $D_{1}(2420)$ mesons decay only via $D^{\ast}\pi$. Both $D\pi$ and
$D^{\ast}\pi$ channels are opened for $D_{2}^{\ast}(2460)$.

In the case of BaBar data~\cite{PhysRevLett.101.261802,PhysRevLett.103.051803} the
branching fractions $\mathcal{B}(D_{2}^{\ast}(2460) \to D^{(\ast)}\pi)$ include both
the $D^{\ast}$ and $D$ contributions. As they also provide the ratio
$\mathcal{B}_{D/D^{(\ast)}}$, we estimate separately the $D^{\ast}$ and $D$
contributions.

A similar analysis can be done in the charmed strange sector for the $B_{s}$ meson
semileptonic decays. Here the intermediate states are the orbitally charmed-strange
mesons, $D_{s}^{\ast\ast}$, and the available final channels are $DK$ and
$D^{\ast}K$. The PDG only reports the value of the following product of branching
fractions ${\cal B}(B_{s}^{0} \to D_{s1}(2536)^{-} \mu^{+}\nu_{\mu}) {\cal
B}(D_{s1}(2536)^{-}\to D^{\ast-}\bar{K}^{0})=2.4\pm0.7$~\cite{PDG2012} based on their
best value for ${\cal B}(\bar{b} \to B_{s}^{0})$ and the experimental data for ${\cal
B}(\bar{b} \to B_{s}^{0}){\cal B}(B_{s}^{0} \to D_{s1}(2536)^{-}
\mu^{+}\nu_{\mu}){\cal B}(D_{s1}(2536)^{-} \to D^{\ast-}\bar{K}^{0})$ measured by the
D0 Collaboration~\cite{PhysRevLett.102.051801}.

All these magnitudes can be consistently calculated in the framework of constituent
quark models because they can simultaneously account for the hadronic part of the
weak process and the strong meson decays. In this context, meson strong decays will
be described through the $^{3}P_{0}$ model presented before. As for the weak process
the matrix elements factorize into a leptonic and a hadronic part. It is the hadronic
part that contains the nonperturbative strong interaction effects and we will
evaluate it within our constituent quark model. Further details on the semileptonic
decay calculation can be found on
Refs.~\refcite{PhysRevD.73.054024,PhysRevD.74.074008,PhysRevD.84.094029}.

The mesons involved in the reactions have been discussed in previous sections of this
work. The most relevant features to take into account here are: we have reached a
good description of the singlet- and triplet-spin $S$-wave charmed mesons, $D$ and 
$D^{\ast}$, and charmed-strange mesons, $D_{s}$ and $D_{s}^{\ast}$. We have seen that
the interpretation of the $D_{s0}^{\ast}(2317)$ as a canonical $c\bar{s}$ state is
plausible since its mass goes down to the experimental value when the one-loop QCD
corrections to the OGE potential are taken into account. The $D_{s1}(2460)$ meson has
an important non-$q\bar{q}$ contribution. The presence of non-$q\bar{q}$ degrees of
freedom in the $J^{P}=1^{+}$ charmed-strange meson sector enhances the $j_{q}=3/2$
component of the $D_{s1}(2536)$ meson, which is almost a pure $q\bar{q}$ state.
Table~\ref{tab:mixedstates} shows only the $q\bar{q}$ probabilities of the orbitally
excited charmed and charmed strange mesons.

\begin{table}[t!]
\begin{center}
\caption{\label{tab:mixedstates} Probability distributions and their relative phases
for the four states predicted by CQM. In the $1^{+}$ strange sector the effects of
non-$q\bar q$ components are included; see text for details.}
\begin{tabular}{ccccc}
\toprule
& $D_{0}^{\ast}(2400)$ & $D_{1}(2420)$ & $D_{1}(2430)$ & $D_{2}^{\ast}(2460)$ \\
\colrule
$^{3}P_{0}$ & $+,\,1.0000$ & - & - & - \\
$^{1}P_{1}$ & - & $-,\,0.5903$ & $-,\,0.4097$ & - \\
$^{3}P_{1}$ & - & $+,\,0.4097$ & $-,\,0.5903$ & - \\
$^{3}P_{2}$ & - & - & - & $+,\,0.99993$ \\
\colrule
$1/2,0^+$ & $+,\,1.0000$ & - & - & - \\
$1/2,1^+$ & - & $+,\,0.0063$ & $-,\,0.9937$ & - \\
$3/2,1^+$ & - & $+,\,0.9937$ & $+,\,0.0063$ & - \\
$3/2,2^+$ & - & - & - & $+,\,0.99993$ \\
\toprule
& $D_{s0}^{\ast}(2317)$ & $D_{s1}(2536)$ & $D_{s1}(2460)$ &
$D_{s2}^{\ast}(2573)$ \\
\colrule
$^{3}P_{0}$ & $+,\,1.0000$ & - & - & - \\
$^{1}P_{1}$ & - & $-,\,0.7210$ & $-,\,0.1880$ & - \\
$^{3}P_{1}$ & - & $+,\,0.2770$ & $-,\,0.5570$ & - \\
$^{3}P_{2}$ & - & - & - & $+,\,0.99991$ \\
\colrule
$1/2,0^+$ & $+,\,1.0000$ & - & - & - \\
$1/2,1^+$ & - & $-,\,0.0038$ & $-,\,0.7390$ & - \\
$3/2,1^+$ & - & $+,\,0.9942$ & $-,\,0.0060$ & - \\
$3/2,2^+$ & - & - & - & $+,\,0.99991$ \\
\botrule
\end{tabular}
\end{center}
\end{table}

\begin{table}[t!]
\begin{center}
\caption{\label{tab:experiment2} Most recent experimental measurements reported by
the Belle and BaBar Collaborations and their comparison with our results. $l$ stands
for a light $e$ or $\mu$ lepton. The symbol $(\ast)$ indicates the estimated results
from the original data using $B_{D/D^{(\ast)}}$.}
\scalebox{0.80}{\begin{tabular}{lccc}
\toprule
& Belle~\cite{PhysRevD.77.091503} &
BaBar~\cite{PhysRevLett.101.261802,PhysRevLett.103.051803} & Theory \\
& $(\times10^{-3})$ & $(\times10^{-3})$ & $(\times10^{-3})$ \\
\colrule
$D_{0}^{\ast}(2400)$ & & & \\[2ex]
${\cal B}(B^+ \to \bar{D}^{\ast}_{0}(2400)^{0} l^+ \nu_l){\cal
B}(\bar{D}^{\ast}_{0}(2400)^{0}\to D^{-}\pi^+)$ & $2.4\pm0.4\pm0.6$ &
$2.6\pm0.5\pm0.4$ & $2.15$ \\
${\cal B}(B^0 \to D^{\ast}_{0}(2400)^{-} l^+ \nu_l){\cal
B}(D^{\ast}_{0}(2400)^{-}\to \bar{D}^{0}\pi^-)$ & $2.0\pm0.7\pm0.5$ &
$4.4\pm0.8\pm0.6$ & $1.80$ \\
\colrule
$D_{1}(2430)$ & & & \\[2ex]
${\cal B}(B^+ \to \bar{D}_{1}(2430)^{0} l^+ \nu_l){\cal
B}(\bar{D}_{1}(2430)^{0}\to D^{\ast-}\pi^+)$ & $<0.7$ & $2.7\pm0.4\pm0.5$ &
$1.32$ \\
${\cal B}(B^0 \to D_{1}(2430)^{-} l^+ \nu_l){\cal B}(D_1(2430)^{-}\to
\bar{D}^{\ast0}\pi^-)$ & $<5$ & $3.1\pm0.7\pm0.5$ & $1.23$ \\
\colrule
$D_{1}(2420)$ & & & \\[2ex]
${\cal B}(B^+ \to \bar{D}_1(2420)^{0} l^+ \nu_l){\cal B}(\bar{D}_1(2420)^{0}\to
D^{\ast-}\pi^+)$ & $4.2\pm0.7\pm0.7$ & $2.97\pm0.17\pm0.17$ & $2.57$ \\
${\cal B}(B^0 \to D_1(2420)^{-} l^+ \nu_l){\cal B}(D_1(2420)^{-}\to
\bar{D}^{\ast0}\pi^-)$ & $5.4\pm1.9\pm0.9$ & $2.78\pm0.24\pm0.25$ & $2.39$ \\
\colrule
$D_{2}^{\ast}(2460)$ & & & \\[2ex]
${\cal B}(B^+ \to \bar{D}^{\ast}_2(2460)^{0} l^+ \nu_l){\cal
B}(\bar{D}^{\ast}_2(2460)^{0}\to D^{-}\pi^+)$ & $2.2\pm0.3\pm0.4$ &
$1.4\pm0.2\pm0.2^{(\ast)}$ & $1.43$ \\
${\cal B}(B^+ \to \bar{D}^{\ast}_2(2460)^{0} l^+ \nu_l){\cal
B}(\bar{D}^{\ast}_2(2460)^{0}\to D^{\ast-}\pi^+)$ & $1.8\pm0.6\pm0.3$ &
$0.9\pm0.2\pm0.2^{(\ast)}$ & $0.79$ \\
${\cal B}(B^+ \to \bar{D}^{\ast}_2(2460)^{0} l^+ \nu_l){\cal
B}(\bar{D}^{\ast}_2(2460)^{0}\to D^{(\ast)-}\pi^+)$ & $4.0\pm0.7\pm0.5$ &
$2.3\pm0.2\pm0.2$ & $2.22$ \\[2ex]
${\cal B}(B^0 \to D^{\ast}_2(2460)^{-} l^+ \nu_l){\cal
B}(D^{\ast}_2(2460)^{-}\to \bar{D}^{0}\pi^-)$ & $2.2\pm0.4\pm0.4$ &
$1.1\pm0.2\pm0.1^{(\ast)}$ & $1.34$ \\
${\cal B}(B^0 \to D^{\ast}_2(2460)^{-} l^+ \nu_l){\cal
B}(D^{\ast}_2(2460)^{-}\to \bar{D}^{\ast0}\pi^-)$ & $<3$ &
$0.7\pm0.2\pm0.1^{(\ast)}$ & $0.74$ \\
${\cal B}(B^0 \to D^{\ast}_2(2460)^{-} l^+ \nu_l){\cal
B}(D^{\ast}_2(2460)^{-}\to \bar{D}^{(\ast)0}\pi^-)$ & $<5.2$ & $1.8\pm0.3\pm0.1$
& $2.08$ \\[2ex]
${\cal B}_{D/D^{(\ast)}}$ & $0.55\pm0.03$ & $0.62\pm0.03\pm0.02$ & $0.65$ \\
\botrule
\end{tabular}}
\end{center}
\end{table}

Table~\ref{tab:experiment2} shows the final results and their comparison
with the experimental data in the case of the $B$ semileptonic decays into orbitally
excited charmed mesons.

The meson $D_{0}^{\ast}(2400)$ has $J^{P}=0^{+}$ quantum numbers and, therefore, due
to parity conservation, it decays only into $D\pi$, so that we have ${\cal
B}(\bar{D}^{\ast}_{0}(2400)^{0}\to D^{-}\pi^+)={\cal B}(D^{\ast}_{0}(2400)^{-} \to
\bar{D}^{0}\pi^-)=2/3$ coming from isospin symmetry. One can see in
Table~\ref{tab:experiment2} that the theoretical product of branching fractions agrees
well with the latest BaBar data. The difference between the semileptonic width of the
charged and neutral $B$ mesons is due to the large mass difference between the
$D_{0}^{\ast}(2400)^{0}$ and $D_{0}^{\ast}(2400)^{\pm}$ mesons for which we take the
masses reported in Ref.~\refcite{PDG2012}.

The only OZI-allowed decay channel for the $D_{1}(2430)$ meson is the $D_{1}(2430)\to
D^{\ast}\pi$ so that isospin symmetry predicts a branching fraction
$\mathcal{B}(D_{1}(2430)\to D^{\ast}\pi^{\pm})=2/3$. We have in this case for the product of
branching fractions shown in Table~\ref{tab:experiment2} that our value is roughly a factor of $2$ smaller than the results from the
BaBar Collaboration~\cite{PhysRevLett.101.261802}. Note however the disagreement
between BaBar and Belle data for the product of branching fractions in which the
$\bar{D}_{1}(2430)^{0}$ meson is involved.

As in the previous case, the branching fraction $\mathcal{B}(D_{1}(2420) \to
D^{\ast}\pi^{\pm})$ is again $2/3$ in our model because $D_{1}(2420)\to
D^{\ast}\pi$ is the only OZI-allowed decay channel. The products of branching
fractions compare very well with the latest BaBar data~\cite{PhysRevLett.103.051803},
as seen in Table~\ref{tab:experiment2}.

The strong decays which appear in the decay chains that involve the
$D_{2}^{\ast}(2460)$ meson are $D_{2}^{\ast}(2460) \to D\pi^{-}$ and
$D_{2}^{\ast}(2460)\to D^{\ast}\pi^{-}$. In Table~\ref{tab:ratiosD2} we show the
strong decay branching ratios obtained with the $^{3}P_{0}$ model. They are in good
agreement with experimental data~\cite{PDG2012}. 
Considering that the total width of the $D_{2}^{\ast}(2460)$ meson is the
sum of the partial widths of $D\pi$ and $D^{\ast}\pi$ channels, since these are
the only OZI-allowed processes, we are able to predict the
products of branching fractions in Table~\ref{tab:experiment2} which are
in very good agreement with BaBar
data~\cite{PhysRevLett.103.051803}.

\begin{table}[t!]
\begin{center}
\caption{\label{tab:ratiosD2} Open-flavor strong branching ratios for
$D_{2}^{\ast}(2460)$ meson calculated through the $^{3}P_{0}$ model and their
comparison with those collected by the PDG~\cite{PDG2012}.}
\begin{tabular}{lcc}
\toprule
Branching ratio & Theory & Experiment \\
\colrule
$\Gamma(D_{2}^{\ast+} \to D^{0}\pi^{+})/\Gamma(D_{2}^{\ast+} \to
D^{\ast0}\pi^{+})$ & $1.80$ & $1.9\pm1.1\pm0.3$ \\
$\Gamma(D_{2}^{\ast0} \to D^{+}\pi^{-})/\Gamma(D_{2}^{\ast0} \to
D^{\ast+}\pi^{-})$ & $1.82$ & $1.56\pm0.16$ \\
$\Gamma(D_{2}^{\ast} \to D\pi)/\Gamma(D_{2}^{\ast} \to D^{(\ast)}\pi)$ & $0.65$ &
$0.62\pm0.03\pm0.02$ \\
\botrule
\end{tabular}
\end{center}
\end{table}

The semileptonic decays of $B_{s}$ meson into orbitally excited charmed-strange
mesons ($D_{s}^{\ast\ast}$) provides an extra opportunity to get more insight into
the structure of these latter mesons.

\begin{figure}[t!]
\begin{center}
\parbox[c]{0.49\textwidth}{
\centering
\includegraphics[width=0.40\textwidth]{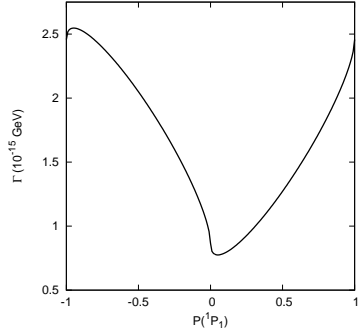}
}
\caption{\label{fig:resDs12460} Decay width for the $B_{s}^{0} \to
D_{s1}(2460)^{-}\mu^{+}\nu_{\mu}$ decay as a function of the $^{1}P_{1}$ component
probability. The sign reflects the relative phase between $^{1}P_{1}$ and $^{3}P_{1}$
components: $-1$ opposite phase and $+1$ same phase.}
\end{center}
\end{figure}

We have calculated the semileptonic $B_{s}$ decays assuming that the
$D_{s}^{\ast\ast}$ mesons are pure $q\bar{q}$ systems. For the $D_{s0}^\ast(2317)$
and $D_{s1}(2460)$, which are below the corresponding $DK$ and $D^{\ast}K$
thresholds, we only quote the weak decay branching fractions. Concerning the
$D_{s1}(2460)$, the $^{1}P_{1}$ and $^{3}P_{1}$ probabilities change with the
coupling to non-$q\bar{q}$ degrees of freedom. What we do here is to vary these
probabilities (including the phase) in order to obtain the limits of the decay width
in the case of the $D_{s1}(2460)$ being a pure $q\bar{q}$ state, see
Fig.~\ref{fig:resDs12460}. Assuming that non-$q\bar{q}$ components will give a small
contribution to the weak decay, experimental results lower than these limits will be
an indication of a more complex structure for this meson.

For the decay into $D_{s1}(2536)$, our model predicts the weak decay branching
fraction ${\cal B}(B_s^0\to D_{s1}(2536)\mu^+ \nu_\mu)=4.77\times 10^{-3}$ and
the strong branching fraction ${\cal B}(D_{s1}(2536)^{-}\to
D^{\ast-}\bar{K}^{0}) = 0.43$. The final result appears in
Table~\ref{tab:experiment3}. It is compatible with the  existing experimental
data~\cite{PDG2012}, which to us is a confirmation of our result about the $q\bar{q}$
nature of this state.

\begin{table}[t!]
\begin{center}
\caption{\label{tab:experiment3} Our predictions and their comparison with the
available experimental data for semileptonic $B_{s}$ decays into orbitally excited
charmed-strange mesons.}
\scalebox{0.90}{\begin{tabular}{lcc}
\toprule
& Experiment & Theory \\
& $(\times10^{-3})$ & $(\times10^{-3})$ \\
\colrule
$D_{s0}^{\ast}(2317)$ & & \\[2ex]
${\cal B}(B_{s}^{0} \to D^{\ast}_{s0}(2318)^{-} \mu^+ \nu_\mu)$ & - & $4.43$ \\
\colrule
$D_{s1}(2460)$ & & \\[2ex]
${\cal B}(B_{s}^{0} \to D_{s1}(2460)^{-} \mu^+ \nu_\mu)$ & - & $1.74-5.70$ \\
\colrule
$D_{s1}(2536)$ & & \\[2ex]
${\cal B}(B_{s}^{0} \to D_{s1}(2536)^{-} \mu^+ \nu_\mu ){\cal
B}(D_{s1}(2536)^{-}\to D^{\ast-}\bar{K}^{0})$ &
$2.4\pm0.7$~\cite{PDG2012,PhysRevLett.102.051801} & $2.05$ \\
\colrule
$D_{s2}^{\ast}(2573)$ & & \\[2ex]
${\cal B}(B_{s}^0 \to D^{\ast}_{s2}(2573)^{-} \mu^+ \nu_\mu){\cal
B}(D^{\ast}_{s2}(2573)^{-} \to D^{-}\bar{K}^{0})$ & - & $1.70$ \\
${\cal B}(B_{s}^0 \to D^{\ast}_{s2}(2573)^{-} \mu^+ \nu_\mu){\cal
B}(D^{\ast}_{s2}(2573)^{-} \to D^{\ast-}\bar{K}^{0})$ & - & $0.18$ \\
${\cal B}(B_{s}^0 \to D^{\ast}_{s2}(2573)^{-} \mu^+ \nu_\mu){\cal
B}(D^{\ast}_{s2}(2573)^{-} \to D^{(\ast)-}\bar{K}^{0})$ & - & $1.88$ \\
\botrule
\end{tabular}}
\end{center}
\end{table}

In the case of the $D_{s2}^{\ast}(2573)$ meson the open strong decays are $DK$
and $D^{\ast}K$, so the experimental measurements must be referred to ${\cal
B}(B_{s}^0 \to D^{\ast}_{s2}(2573)^{-} \mu^+ \nu_\mu)$ ${\cal
B}(D^{\ast}_{s2}(2573)^{-} \to D^{-}\bar{K}^{0})$ and ${\cal B}(B_{s}^0 \to
D^{\ast}_{s2}(2573)^{-} \mu^+ \nu_\mu){\cal B}(D^{\ast}_{s2}(2573)^{-} \to
D^{\ast-}\bar{K}^{0})$. For the weak branching fraction we get in this case ${\cal
B}(B_{s}^{0} \to D^{\ast}_{s2}(2573)^{-} \mu^+\nu_\mu)=3.76\times10^{-3}$. For the
strong decay part of the reaction, we obtain
\begin{equation}
\begin{split}
{\cal B}(D_{s2}^{\ast-}\to D^{-}\bar{K}^{0}) &= 0.45 \\
{\cal B}(D_{s2}^{\ast-}\to D^{\ast-}\bar{K}^{0}) &= 0.047 \\
\end{split}
\end{equation}
using the $^{3}P_{0}$ model. Besides we predict the ratio
\begin{equation}
\frac{\Gamma(D_{s2}^{\ast}\to DK)}{\Gamma(D_{s2}^{\ast}\to
DK)+\Gamma(D_{s2}^{\ast}\to D^{\ast}K)} = 0.91.
\end{equation}
Our final results can be seen in Table.~\ref{tab:experiment3}.

\subsection{Nonleptonic $B$ decays into $D^{(\ast)}D_{sJ}$ final states}

The nonleptonic decays of $B$ mesons have been used to search for new charmonium and
charmed-strange mesons and to study their properties in detail. For instance, the
properties of the $D_{s0}^{\ast}(2317)$ and $D_{s1}(2460)$ mesons were not well known
until the Belle Collaboration observed the $B \to \bar{D}D_{s0}^{\ast}(2317)$ and $B
\to \bar{D}D_{s1}(2460)$ decays~\cite{PhysRevLett.91.262002}.

First observations of the $B \to \bar{D}^{(\ast)}D_{s1}(2536)$ decay modes have been
reported by BaBar~\cite{PhysRevD.74.091101,PhysRevD.77.011102} and an upper limit on
the decay $B^{0} \to D^{\ast-}D_{s1}(2536)^{+}$ was also obtained by
Belle~\cite{PhysRevD.76.072004}. The most recent analysis of the production of
$D_{s1}(2536)^{+}$ in double charmed $B$ meson decays has been reported by the Belle
Collaboration in Ref.~\refcite{PhysRevD.83.051102}. Using the latest measurements of
the $B \to D^{(\ast)}D_{sJ}$ branching fractions~\cite{PDG2012}, they calculated the
ratios
\begin{equation}
\begin{split}
R_{D0} &= \frac{{\cal B}(B \to DD_{s0}^{\ast}(2317))}{{\cal B}(B \to
DD_{s})} = 0.10\pm0.03, \\
R_{D^{\ast}0} &= \frac{{\cal B}(B \to D^{\ast}D_{s0}^{\ast}(2317))}{{\cal B}(B
\to D^{\ast}D_{s})} = 0.15\pm0.06, \\
R_{D1} &= \frac{{\cal B}(B \to DD_{s1}(2460))}{{\cal B}(B \to
DD_{s}^{\ast})} = 0.44\pm0.11, \\
R_{D^{\ast}1} &= \frac{{\cal B}(B \to D^{\ast}D_{s1}(2460))}{{\cal B}(B \to
D^{\ast}D_{s}^{\ast})} = 0.58\pm0.12.
\label{eq:DDsratio1}
\end{split}
\end{equation}
In addition, the same ratios were calculated for $B \to D^{(\ast)}D_{s1}(2536)^{+}$
decays using combined results by the BaBar~\cite{PhysRevD.77.011102} and
Belle~\cite{PhysRevD.83.051102} Collaborations
\begin{equation}
\begin{split}
R_{D1'} &= \frac{{\cal B}(B \to DD_{s1}(2536))}{{\cal B}(B \to
DD_{s}^{\ast})} = 0.049\pm0.010, \\
R_{D^{\ast}1'} &= \frac{{\cal B}(B \to D^{\ast}D_{s1}(2536))}{{\cal B}(B \to
D^{\ast}D_{s}^{\ast})} = 0.044\pm0.010.
\label{eq:DDsratio2}
\end{split}
\end{equation}

From a theoretical point of view, this kind of decays can be described using the
factorization approximation~\cite{PhysRevD.73.054024}. This amounts to evaluate the
matrix element which describes the $B \to D^{(\ast)}D_{sJ}$ weak decay process as a
product of two matrix elements, the first one to describe the $B$ weak transition
into the $D^{(\ast)}$ meson and the second one for the weak creation of the
$c\bar{s}$ pair which makes the $D_{sJ}$ meson. The latter matrix element is
proportional to the corresponding $D_{sJ}$ meson decay constant.

The $D_{sJ}$ meson decay constants are not known experimentally except for the ground
state, $D_{s}$, which has been measured by different Collaborations. Another way to
study $D_{sJ}$ mesons, that does not rely on the knowledge of their decay constants,
is through the decays $B_{s}\to D_{sJ}M$, where $M$ is a meson with a well known
decay constant. However, the experimental study of these processes is currently
difficult for several reasons. First, $B$-factories would need to collect data at the
$\Upsilon(5S)$ resonance. Second, the kinematically clean environment of $B$ meson
decays does not hold in $B_{s}$ decays. And finally, the fraction of events with a
pair of $B_{s}$ mesons over the total number of events with a pair of $b$-flavored
hadrons has been measured to be relatively small, $f_{s}[\Upsilon(5S)] = 0.193 \pm
0.029$. These difficulties leave, for the time being, the  $B \to D^{(\ast)}D_{sJ}$
decay processes as our best option to study $D_{sJ}$ meson properties.

According to Refs.~\refcite{LeYaouanc1996582} and~\refcite{Datta2003164}, within the factorization
approximation and in the heavy quark limit, the ratios in Eqs.~(\ref{eq:DDsratio1})
and~(\ref{eq:DDsratio2}) can be written as
\begin{equation} 
\begin{split}
R_{D0} &= R_{D^{\ast}0} = \left|
\frac{f_{D_{s0}^{\ast}(2317)}}{f_{D_{s}}}\right|^{2}, \\
R_{D1} &= R_{D^{\ast}1} = \left|
\frac{f_{D_{s1}(2460)}}{f_{D_{s}^{\ast}}}\right|^{2}, \\
R_{D1'} &= R_{D^{\ast}1'} = \left|
\frac{f_{D_{s1}(2536)}}{f_{D_{s}^{\ast}}}\right|^{2}, \\
\end{split}
\end{equation}
where the phase space effects are neglected because they are subleading in the
heavy quark expansion. Now, in the heavy quark limit one has
$f_{D_{s0}^{\ast}(2317)}=f_{D_{s1}(2460)}$, $f_{D_{s}} = f_{D_{s}^{\ast}}$ and
$f_{D_{s1}(2536)}=0$. Moreover, there are several estimates of the decay
constants, always in the heavy quark
limit~\cite{Yaouanc200159,Colangelo2002193,PhysRevD.60.033002}, that predict for
$P$-wave, $j_{q}=1/2$ states similar decay constants as for the ground state mesons
(i.e. $f_{D_{s0}^{\ast}(2317)}=f_{D_{s}}$ and $f_{D_{s1}(2460)}=f_{D_{s}^{\ast}}$),
and very small decay constants for $P$-wave, $j_{q}=3/2$ states. 
Thus, in the heavy quark limit one would expect $R_{D0}\sim 1$, $R_{D1}\sim 1$ and $R_{D1'}\ll 1$.
While the decay into $D_{s1}(2536)$ follows the
expectations, this is not the case for the $D_{s0}^{\ast}(2317)$ and $D_{s1}(2460)$
mesons. This fact has motivated to argue that either those two states are not
canonical $c\bar{s}$ mesons or that the factorization approximation does not hold for
decays involving those particles.

Leaving aside that the factorization approximation has been recently analyzed in
Refs.~\refcite{PhysRevD.64.094001,PhysRevD.58.014007,PhysRevD.55.6780} finding that
it works well in this kind of processes, we will concentrate in the influence of the
effect of the finite $c$-quark mass in the theoretical predictions. As found in
Ref.~\refcite{PhysRevD.72.094010}, $1/m_{Q}$ contributions give large corrections to
various quantities describing $B\to D^{\ast\ast}$ transitions and we expect they also
play an important role in this case. It is possible that taking into account the
finite mass of the charmed quark one can distinguish better between $q\bar{q}$ and
non-$q\bar{q}$ structures for the $D_{sJ}$ mesons.

The nonleptonic decay width for $B \to D^{(\ast)}D_{sJ}$ processes in the
factorization approximation and using helicity
formalism~\cite{PhysRevD.73.054024,PhysRevD.74.074008} is given in
Ref.~\refcite{PhysRevD.86.014010}. Using experimental masses we obtain the ratios
\begin{equation}
\begin{split}
R_{D0} &= 0.90 \times \left| \frac{f_{D_{s0}^{\ast}(2317)}}{f_{D_{s}}}
\right|^{2}, \\
R_{D^{\ast}0} &= 0.72 \times \left| \frac{f_{D_{s0}^{\ast}(2317)}}{f_{D_{s}}}
\right|^{2}.
\label{eq:ratiosDs0}
\end{split}
\end{equation}
The double ratio $R_{D^{\ast}0}/R_{D0}$ does not depend on decay constants, and
in our model we obtain $R_{D^{\ast}0}/R_{D0}=0.80$. The experimental value is
given by $R_{D^{\ast}0}/R_{D0}=1.50\pm0.75$. Our result is small compared to the
central experimental value but we are compatible within $1\sigma$. In the case
of the meson $D_{s1}(2460)$ we obtain
\begin{equation}
\begin{split}
R_{D1} &= 0.70 \times \left| \frac{f_{D_{s1}(2460)}}{f_{D_{s}^{\ast}}}
\right|^{2}, \\
R_{D^{\ast}1} &= 1.00 \times \left| \frac{f_{D_{s1}(2460)}}{f_{D_{s}^{\ast}}}
\right|^{2},
\label{eq:ratiosDs1}
\end{split}
\end{equation}
and for the double ratio $R_{D^{\ast}1}/R_{D1}$ we get $1.43$, which agrees
well with the experimental result $R_{D^{\ast}1}/R_{D1}=1.32\pm0.43$. Finally,
for the meson $D_{s1}(2536)$ we obtain
\begin{equation}
\begin{split}
R_{D1'} &= 0.64 \times \left| \frac{f_{D_{s1}(2536)}}{f_{D_{s}^{\ast}}}
\right|^{2}, \\
R_{D^{\ast}1'} &= 0.99 \times \left| \frac{f_{D_{s1}(2536)}}{f_{D_{s}^{\ast}}}
\right|^{2},
\label{eq:ratiosDs1p}
\end{split} 
\end{equation}
and for the double ratio $R_{D^{\ast}1'}/R_{D1'}$, our value is $1.56$ which
in this case is $2\sigma$ above the experimental one, $0.90\pm0.27$. 

The quality of the experimental numbers does not allow to be very conclusive as
to the goodness of the factorization approximation. But one can conclude from
Eqs.~(\ref{eq:ratiosDs0}),~(\ref{eq:ratiosDs1}) and~(\ref{eq:ratiosDs1p}) that the
phase-space and weak matrix element corrections cannot be ignored, as done when using
the infinite heavy quark mass limit.

\begin{table}[!t]
\begin{center}
\caption{\label{tab:DCmodels} Theoretical predictions of decay constants for
pseudoscalar and vector charmed mesons. The data have been taken from
Ref.~\protect\refcite{PDG2012} for pseudoscalar mesons and from
Ref.~\protect\refcite{GuoLi2006492}
for vector mesons. PQL$\,\equiv\,$Partially-Quenched Lattice calculation,
QL$\,\equiv\,$Quenched Lattice calculations, RBS$\,\equiv\,$Relativistic
Bethe-Salpeter, RQM$\,\equiv\,$Relativistic Quark Model,
BS$\,\equiv\,$Bethe-Salpeter Method and RM$\,\equiv\,$Relativistic Mock
meson model. See the text for the meaning of 
symbols $\dag$ and $\ddag$.}
\scalebox{0.80}{\begin{tabular}{llll} 
\toprule
Approach & $f_{D}$ (MeV) & $f_{D_{s}}$ (MeV) & $f_{D_{s}}/f_{D}$ \\[0.2ex]
\colrule
Ours & $297.019^{(\dag)}$ & $416.827^{(\dag)}$ & $1.40^{(\dag)}$ \\
     & $214.613^{(\ddag)}$ & $286.382^{(\ddag)}$ & $1.33^{(\ddag)}$ \\[1.5ex]
Experiment & $206.7\pm8.9$ & $257.5\pm6.1$ & $1.25\pm0.06$ \\
Lattice (HPQCD+UKQCD) & $208\pm4$ & $241\pm3$ & $1.162\pm0.009$ \\
Lattice (FNAL+MILC+HPQCD) & $217\pm10$ & $260\pm10$ & $1.20\pm0.02$ \\
PQL & $197\pm9$ & $244\pm8$ & $1.24\pm0.03$ \\
QL (QCDSF) & $206\pm6\pm3\pm22$ & $220\pm6\pm5\pm11$ & $1.07\pm0.02\pm0.02$ \\
QL (Taiwan) & $235\pm8\pm14$ & $266\pm10\pm18$ & $1.13\pm0.03\pm0.05$ \\
QL (UKQCD) & $210\pm10^{+17}_{-16}$ & $236\pm8^{+17}_{-14}$ &
$1.13\pm0.02^{+0.04}_{-0.02}$ \\
QL & $211\pm14^{+2}_{-12}$ & $231\pm 12^{+6}_{-1}$ & $1.10\pm0.02$\\
QCD Sum Rules & $177\pm21$ & $205\pm22$ & $1.16\pm0.01\pm0.03$ \\
QCD Sum Rules & $203\pm20$ & $235\pm 24$ & $1.15\pm 0.04$ \\
Field Correlators & $210\pm10$ & $260\pm10$ & $1.24\pm0.03$ \\
Light Front & $206$ & $268.3\pm19.1$ & $1.30\pm0.04$ \\
\toprule
Approach & $f_{D^{\ast}}$ (MeV) & $f_{D_{s}^{\ast}}$ (MeV) &
$f_{D_{s}^{\ast}}/f_{D^{\ast}}$ \\[0.2ex]
\colrule
Ours & $247.865^{(\dag)}$ & $329.441^{(\dag)}$ & $1.33^{(\dag)}$ \\[1.5ex]
RBS & $340\pm22$ & $375\pm24$ & $1.10\pm0.06$ \\
RQM & $315$ & $335$ & $1.06$ \\
QL (Italy) & $234$ & $254$ & $1.04\pm0.01^{+2}_{-4}$ \\
QL (UKQCD) & $245\pm20^{+0}_{-2}$ & $272\pm16^{+0}_{-20}$ & $1.11\pm0.03$ \\
BS & $237$ & $242$ & $1.02$ \\
RM & $262\pm10$ & $298\pm11$ & $1.14\pm0.09$ \\
\botrule
\end{tabular}}
\end{center}
\end{table}

The decay constants of pseudoscalar and vector mesons in charmed and charmed-strange
sectors are given in Table~\ref{tab:DCmodels}. We compare our results with the
experimental data and those predicted by different approaches and collected in
Refs.~\refcite{PDG2012,GuoLi2006492}. Our original values are those with the symbol
$(\dag)$. The decay constants of vector mesons agree with other approaches. In the
case of the pseudoscalar mesons, the decay constants are simply too large. The reason
for that is the following: Our CQM presents an OGE potential which has a spin-spin
contact hyperfine interaction that is proportional to a Dirac delta function,
conveniently regularized, at the origin. The corresponding regularization parameter
was fitted to determine the hyperfine splittings between the $n^{1}S_{0}$ and
$n^{3}S_{1}$ states in the different flavor sectors, achieving a good agreement in
all of them. While most of the physical observables are insensitive to the
regularization of this delta term, those related with annihilation processes are
affected. The effect is very small in the $^{3}S_{1}$ channel as the delta term is
repulsive in this case. It is negligible for higher partial waves due to the
centrifugal barrier. However, it is sizable in the $^{1}S_{0}$
channel for which the  delta term is attractive. 

One expects that the wave functions of the $1^{1}S_{0}$ and $1^{3}S_{1}$ states are
very similar~\cite{MB2008455}. In fact, they are equal if the Dirac delta term is
ignored. The values with the symbol $(\ddag)$ in Table~\ref{tab:DCmodels} are
referred to the pseudoscalar decay constants which have been calculated using the
wave function of the corresponding $^{3}S_{1}$ state. We recover the agreement with
experiment and also with the predictions of different theoretical approaches. The
$f_{D_{s}}/f_{D}$ and $f_{D_{s}^{\ast}}/f_{D^{\ast}}$ ratios are also shown in the
last column of Table~\ref{tab:DCmodels}. They are not very sensitive to the  delta
term and our values agree nicely with experiment and the values obtained in other
approaches.

\begin{table}[t!]
\begin{center}
\caption{\label{tab:decayconstants} Decay constants calculated within the CQM
including one-loop QCD corrections to the OGE potential and a non-$q\bar{q}$
structure in channel $1^{+}$.}
\begin{tabular}{cccc}
\toprule
Meson & & $f_{D}$ (MeV) & $\sqrt{M_{D}}f_{D}$ $({\rm GeV}^{3/2})$ \\
\colrule
$D_{s0}^{\ast}(2317)$ & & $118.706$ & $0.181$ \\
$D_{s1}(2460)$ & & $165.097$ & $0.259$ \\
$D_{s1}(2536)$ & & $59.176$ & $0.094$ \\
\botrule
\end{tabular}
\end{center}
\end{table}

Table~\ref{tab:decayconstants} summarizes the remaining decay constants needed
for the calculation we are interested in. There, we show the results from the
constituent quark model in which the $1$-loop QCD corrections to the OGE potential
and the presence of non-$q\bar{q}$ degrees of freedom in $J^{P}=1^{+}$
charmed-strange meson sector are included. If one compares $f_{D_{s}}$
($f_{D_{s}^{\ast}}$) to $f_{D_{s0}^{\ast}(2317)}$ ($f_{D_{s1}(2460)}$), one
finds that the latter is suppressed.

Our results for the decay constants clearly deviate from the ones obtained in
the infinite heavy quark mass limit. In that limit one gets
$f_{D_{s0}^{\ast}(2317)} = f_{D_{s}}$, $f_{D_{s1}(2460)}=f_{D_{s}^{\ast}}$ and
$f_{D_{s1}(2536)} = 0$, results that lead to a strong disagreement with
experiment for the decay width ratios in Eqs.~(\ref{eq:DDsratio1})
and~(\ref{eq:DDsratio2}). That was already noticed in Ref.~\refcite{Datta2003164},
where the authors, using the experimental ratios, estimated  that
$f_{D_{s0}^{\ast}(2317)}\sim\frac{1}{3}f_{D_{s}}$ and
$f_{D_{s0}^{\ast}(2317)}\sim f_{D_{s1}(2460)}$ instead. We obtain
$f_{D_{s0}^{\ast}(2317)}/f_{D_{s}}=0.36$, $f_{D_{s0}^{\ast}(2317)} \sim
0.72 f_{D_{s1}(2460)}$ and $f_{D_{s1}(2536)}=59.176\,{\rm MeV}$, the latter
being small compared to the others but certainly not zero.

\begin{table}[t!]
\begin{center}
\caption{\label{tab:Ourratios} Ratios of branching fractions for nonleptonic decays
$B\to D^{(\ast)}D_{sJ}$. The symbol $(\ast)$ indicates that the ratios have been
calculated using the experimental pseudoscalar decay constant in
Table~\ref{tab:DCmodels}. For the $D_{s1}(2460)$ and $D_{s1}(2536)$ mesons, the
ratios have been calculated without $(1)$ and with $(2)$ taking into account the
non-$q\bar{q}$ degrees of freedom in the $J^{P}=1^{+}$ channel.}
\scalebox{0.70}{\begin{tabular}{lcccccc}
\toprule
& \multicolumn{2}{c}{$X\equiv D_{s0}^{\ast}(2317)$} &
\multicolumn{2}{c}{$X\equiv D_{s1}(2460)$} &
\multicolumn{2}{c}{$X\equiv D_{s1}(2536)$} \\
& The. & Exp. & The. & Exp. & The. & Exp. \\
\colrule
${\cal B}(B \to DX)/{\cal B}(B \to DD_{s})$ & $0.19^{(\ast)}$ &
$0.10\pm0.03$ & - & - & - & - \\[2ex]
${\cal B}(B \to D^{\ast}X)/{\cal B}(B \to D^{\ast}D_{s})$
& $0.15^{(\ast)}$ & $0.15\pm0.06$ & - & - & - & - \\[2ex]
${\cal B}(B \to DX)/{\cal B}(B \to DD_{s}^{\ast})$ & - & - &
$\left[\begin{matrix} 0.176^{(1)} \\ 0.177^{(2)} \end{matrix}\right]$ &
$0.44\pm0.11$ & $\left[\begin{matrix} 0.071^{(1)} \\ 0.021^{(2)}
\end{matrix}\right]$ & $0.049\pm0.010$ \\[2ex]
${\cal B}(B \to D^{\ast}X)/{\cal B}(B \to
D^{\ast}D_{s}^{\ast})$ & - & - & $\left[\begin{matrix} 0.251^{(1)} \\
0.252^{(2)} \end{matrix}\right]$ & $0.58\pm0.12$ & $\left[\begin{matrix}
0.110^{(1)} \\ 0.032^{(2)} \end{matrix}\right]$ &
$0.044\pm0.010$ \\
\botrule
\end{tabular}}
\end{center}
\end{table}

Finally, we show in Table~\ref{tab:Ourratios} our results for the ratios written
in Eqs.~(\ref{eq:DDsratio1}) and~(\ref{eq:DDsratio2}). The symbol $(\ast)$
indicates that the ratios have been calculated using the experimental
pseudoscalar decay constant in Table~\ref{tab:DCmodels}. We get results close to
or within the experimental error bars for the $D_{s0}^{\ast}(2317)$ meson, which
to us is an indication that this meson could be a canonical $c\bar{s}$ state.
The incorporation of the non-$q\bar{q}$ degrees of freedom in the $J^{P}=1^{+}$
channel, enhances the $j_{q}=3/2$ component of the $D_{s1}(2536)$ meson and it
gives rise to ratios in better agreement with experiment. Note that the
$D_{s1}(2536)$ meson is an almost pure $q\bar{q}$ state in our description.

The situation is more complicated for the $D_{s1}(2460)$ meson. The probability
distributions of its $^{1}P_{1}$ and $^{3}P_{1}$ components are corrected by the
inclusion of non-$q\bar{q}$ degrees of freedom, the latter making a $\sim25\%$
of the wave function. In our calculation, only the pure $q\bar{q}$ component 
of the $D_{s1}(2460)$ meson has been used to evaluate the $\Gamma(B\to
D^{(\ast)}D_{s1}(2460))$ decay width. The values we get for the corresponding
ratios in Eqs.~(\ref{eq:DDsratio1}) are lower than the experimental data.

%%%%%%%%%%%%%%%%%%%%%%%%%%%%%%%%%%%%%%%%%%%%%%%%%%%%%%%%%%%%%%%%%%%%%%%%%%%%%%%%
\FloatBarrier
%%%%%%%%%%%%%%%%%%%%%%%%%%%%%%%%%%%%%%%%%%%%%%%%%%%%%%%%%%%%%%%%%%%%%%%%%%%%%%%%

\section{Summary}
\label{sec:Summary}

A study of heavy meson properties within a nonrelativistic constituent quark model,
which successfully describes hadron phenomenology and hadronic reactions, has been
presented in this review. Within the heavy quark sector, we have focused on the
spectroscopy and on the electromagnetic, strong and weak processes. 

An exhaustive study of heavy meson spectra in terms of $q\bar{q}$ components has been
performed. The model can be used as a template against which to compare the new 
mesons, whose nature is still unknown and some of them are in conflict with naive
quark model spectations. Some electromagnetic decays have been included. The study
of these processes could provide valuable information on the meson structure
since the operator of electromagnetic transitions is very well known. 

A quite reasonable global description of the charmonium spectra and decay widths has been reached.
One striking feature of our model is the new assignment of the $\psi(4415)$ as a
$D$-wave state leaving the $4S$ state for the $X(4360)$. This agrees with the
last measurements of its leptonic and total decay widths. We have also tested that our
result is compatible with the measured exclusive cross section for the
processes $e^{+}e^{-}\to D^{0}D^{-}\pi^{+}$ and $e^{+}e^{-}\to
D^{0}D^{\ast-}\pi^{+}$. 
Tentative assignments of some $XYZ$ mesons as $c\bar{c}$ states have been done.

The situation is more complicate in the open charm and charmed-strange sectors.
We describe the charmed and charmed-strange mesons $D^{(*)}$ and $D_s^{(*)}$ 
and the $j_{q}^{P}=\frac{3}{2}^{+}$ doublet of the orbitally excited states. 
For the $j_{q}^{P}=\frac{1}{2}^{+}$ doublet the introduction of one loop
correction to the OGE potential brings the mass of the $0^+$ state closer
to experiment but it is not enough to solve the puzzle of these $P$-wave
states. 

We have assumed the presence of non-$q\bar{q}$ degrees of freedom in the
$J^{P}=1^{+}$ charmed-strange meson sector to enhance the $j_{q}=3/2$ component of
the $D_{s1}(2536)$ meson. Independently of the mechanism that produces this effect,
it has become clear that the description of the $D_{s1}(2536)$ meson as a $j_{q}=3/2$
$c\bar{s}$ state is necessary to simultaneously explain its decay
properties. The $J^{P}=1^{+}$ $D_{s1}(2460)$ has an important non-$q\bar q$
contribution in our framework.

In the last years several new resonances on the open charm sector have
been observed. We have discussed their possible quantum numbers attending to
their masses and strong decays. We can only partially describe these states,
although the experimental situation is still not clear.

We have also performed a calculation of the branching fractions for the semileptonic
decays of $B$ and $B_{s}$ mesons into final states containing orbitally excited
charmed and charmed-strange mesons, respectively. Our results for $B$ semileptonic
decays into $D_{0}^{\ast}(2400)$, $D_{1}(2420)$ and $D_{2}^{\ast}(2460)$ are in good
agreement with the latest experimental measurements. In the case of the $D_{1}(2430)$
meson, the prediction lies a factor of $2$ below BaBar. In the case of $B_{s}$
semileptonic decays, our prediction for the ${\cal B}(B_{s}^{0} \to D_{s1}(2536)^{-}
\mu^+ \nu_\mu){\cal B}(D_{s1}(2536)^{-}\to D^{\ast-}\bar{K}^{0})$ product of
branching fractions is in agreement with the experimental data. This, together with
the strong decay properties studied for the $D_{s1}(2536)$ meson, is to us evidence
of a dominant $q\bar q$ structure for this state. We have given also
predictions for decays into other $D_{s}^{\ast\ast}$ mesons which can be useful to
test the $q\bar{q}$ nature of these states.

An analysis of the nonleptonic $B$ meson decays into $D^{(\ast)}D_{sJ}$  
has been also included since it provides valuable
information about the structure of the $D_{s0}^{\ast}(2317)$, $D_{s1}(2460)$ and
$D_{s1}(2536)$ mesons. 
The strong disagreement found between the heavy quark limit predictions and the experimental
data is an indication of the finite $c$-quark mass effects, which are included
in the context of the constituent quark model. 
We have got results
close to or within the experimental error bars for the $D_{s0}^{\ast}(2317)$ meson,
which is an indication that this meson could be a canonical $c\bar{s}$ state. The
description of the $D_{s1}(2536)$ meson as an almost $1^{+}$, $j_{q}=3/2$ $c\bar{s}$
state provides theoretical ratios in better agreement with experiment. 

To conclude, we have tried to show in this review that many aspects of the charmonium physics
can be understood within the framework of the constituent quark model. However there
remains interesting open questions which need further theoretical and experimental
effort to be clarified.

%%%%%%%%%%%%%%%%%%%%%%%%%%%%%%%%%%%%%%%%%%%%%%%%%%%%%%%%%%%%%%%%%%%%%%%%%%%%%%%%

\section*{Acknowledgements}

This work has been partially funded by the U.S. Department of Energy, Office of
Nuclear Physics, under contract No. DE-AC02-06CH11357. Ministerio de Ciencia y
Tecnolog\'ia under contract No. FPA2010-21750-C02-02 and  FIS2011-28853-C02-02, by the European
Community-Research Infrastructure Integrating Activity 'Study of Strongly Interacting
Matter' (HadronPhysics3 Grant No. 283286), by the Spanish Ingenio-Consolider 2010
Program CPAN (CSD2007-00042).

%%%%%%%%%%%%%%%%%%%%%%%%%%%%%%%%%%%%%%%%%%%%%%%%%%%%%%%%%%%%%%%%%%%%%%%%%%%%%%%%

\bibliographystyle{ws-ijmpe}

\bibliography{biblIJMPE_v3}

\end{document}